# Effect of hydrogen gas on magnetic properties of alloys of ferromagnetic metals with Pd and its application in hydrogen gas sensing

Ivan S. Maksymov[1] and M. Kostylev[2,*]

[1]Optical Sciences Centre, Swinburne University of Technology, Hawthorn, VIC 3122, Australia; E-mail: imaksymov@swin.edu.au

[2]Department of Physics and Astrophysics, University of Western Australia, Crawley, WA 6009, Australia

[*]Corresponding author, E-mail: mikhail.kostylev@uwa.edu.au

**Abstract**: The mass-production of fuel-cell vehicles and the eventual transition to the hydrogen economy will require safe, inexpensive and reliable sensors capable of simultaneously detecting low concentrations of leaking hydrogen and measuring broad ranges of hydrogen concentration in storage and energy generating systems. Although several competing sensor technologies can potentially be used in this role, just a few of them have thus far demonstrated a combination of all desirable characteristics. This group of devices also includes magneto-electronic sensors that can detect the presence of hydrogen gas in a range of hydrogen concentrations from zero to 100% at atmospheric pressure with the response time approaching the industry standard of one second. The hydrogen gas sensing mechanism underpinning the operation of magneto-electronic sensors is based on the physical processes of ferromagnetic resonance, magneto-optical Kerr effect and anomalous Hall effect that enable one to measure hydrogen-induced changes in the magnetic properties of structures combining Pd with one or several ferromagnetic metals such as Co, Fe or Ni. In this chapter, we overview the physical foundations of emergent ferromagnetic Pd-alloy-based magneto-electronic hydrogen sensors and compare their characteristics with those of high-performing multilayer thin film-based counterparts that have already demonstrated a potential to find commercial applications.

**Nomenclature**

| | |
|---|---|
| AHE | Anomalous Hall effect |
| CPW | Co-planar waveguide |
| DFT | Density-Functional Theory |
| FMR | Ferromagnetic resonance |
| FM | Ferromagnetic metal |
| $H_{PMA}$ | Effective field of perpendicular magnetic anisotropy |
| IP | In-plane |
| LMA | Longitudinal magnetic anisotropy |
| MOKE | Magneto-optical Kerr effect |
| NM | Non-magnetic metal |
| OOP | Out-of-plane |
| PMA | Perpendicular magnetic anisotropy |
| PP | Perpendicular-to-plane |
| PCB | Printed circuit board |
| SQUID | Superconducting quantum interference device |
| STEM/EDX | Energy dispersive X-ray spectroscopy |
| TEM | Transmission electron microscopy |
| XANES | X-ray absorption near edge structure |
| XMCD | X-ray magnetic dichroism |
| XRD | X-ray diffraction |
| XRR | X-ray reflectivity |
| $4\pi M_{eff}$ | Effective magnetisation |
| $\alpha_G$ | Gilbert damping constant |
| $\Delta H$ | Linewidth of the FMR resonance line |
| $\Delta H_0$ | Inhomogeneous linewidth broadening |

# 1. Introduction and motivation

## 1.1. Hydrogen and fuel-cell vehicle technology

Hydrogen is the most common chemical in the universe. Apart from a number of applications as fuel for transport and heating, a means of storage and transportation of renewable energy and as a raw material in industrial processes [1], hydrogen can also be used as the fuel in fuel cells that produce electricity with high efficiency directly on board of a moving vehicle [1]. The emergent fuel-cell technology opens up considerable opportunities for car manufacturers [2] because, unlike fully electric vehicles, hydrogen fuel-cell cars produce the electricity themselves. At present, a commercially-sold hydrogen electric car can travel a distance of more than 500 km on a single tank, which is comparable with the range an average petrol car can reach on a full tank and is larger than the range of a typical electric car on a full single charge. At the same time, whereas the time required to refuel a hydrogen-powered electric car in a dedicated service station has already approached the typical time spent to refuel a petrol car, more than 30 minutes may be needed to recharge the battery of a battery-powered electric vehicle. Significantly, the hydrogen fuel-cell technology may even be more important for truck manufacturers. Indeed, while the weight of Li-ion batteries used in passenger electric cars has not been of major concern, it has been a prohibitive factor in the truck industry, where currently there is an agreement that hydrogen-powered electric powertrains would be the only option for environmentally friendly trucks of the future [3, 4].

## 1.2. Hydrogen gas sensors and concentration meters

However, several technical challenges have thus far prevented the roll-out of a reliable large-scale hydrogen-based technologies, including hydrogen fuel cells. In particular, there have been significant safety concerns due to the high reactivity of hydrogen fuel with environmental oxygen in the air. Indeed, hydrogen is flammable over a very wide range of concentrations in air (4%–75%) and it is explosive over a wide range of concentrations (15%–59%) at a standard atmospheric temperature [5]. This means that, when ignited in an enclosed space, a hydrogen leak will most likely lead to an explosion. As a result, the use of hydrogen becomes especially dangerous in enclosed areas such as tunnels for vehicular road traffic and underground car parkings.

Thus, to enable the use of hydrogen-based technologies in cars, trucks, buses and other vehicles, there is a need in efficient and low-cost hydrogen gas sensors and concentration meters. Sensors suitable for this role must reliably detect even the weakest leakages of hydrogen gas and therefore be susceptible to very small gas concentrations (in the ppm range). Significantly, the response time of a hydrogen sensor must be less than one second [5] to ensure that the concentration of leaking hydrogen does not exceed a dangerous level. On the other hand, hydrogen concentration meters must have a high resolution in the gas concentration range of more than 20%. At the same time, they should not lose their measuring efficiency near the 100% gas concentration mark [5].

These stringent specifications challenge any existing hydrogen sensor technology [5], despite commercial maturity of several kinds of hydrogen sensors that have been employed for decades in various industrial settings [5]. This is because those sensors were designed to operate in specific environments that are different from those of hydrogen fuel cells and relevant technologies. This situation motivates the development of new hydrogen gas sensors intended to meet the demands of the emergent fuel-cell car technology and, more broadly, to support the transition to hydrogen economy (for a review see, e.g., [5, 6, 7, 8, 9, 10]). However, despite significant recent improvements, most of the existing hydrogen gas sensor still suffer from either insufficient sensitivity, slow response time, high power consumption or potential flammability issues [8, 11] and therefore they are not mass production ready. Subsequently, the door remains open for alternative approaches to hydrogen gas sensing.

*1.3. Novel approaches to hydrogen gas sensing*

Recently, several novel methods of hydrogen gas sensing were proposed using a combination of physical phenomena known from the fields of magnetism and magneto-electronics [12, 13, 14, 15, 16]. Those works demonstrated that the resulting magneto-electronic devices open up considerable opportunities for ultra-low fire-hazard, contactless and simple detection of hydrogen in virtually every practical situation.

Magneto-electronic sensors exploit reversible hydrogen-induced changes in magnetic properties of thin magnetic films made of alternating layers of a ferromagnetic metal (FM) such as cobalt (Co), iron (Fe) or nickel (Ni) and non-magnetic metal (NM) palladium (Pd) layers (see, e.g., [12, 13, 14, 17]). A number of magneto-electronic sensor prototypes based on a ferromagnetic resonance (FMR), magneto-optical Kerr effect (MOKE), anomalous Hall effect (AHE) and other detection mechanisms have been tested in a range of hydrogen concentrations from zero to 100% at atmospheric pressure and all of them showed highly promising results (Section 2).

Of course, the application of Pd in a sensitive element of the emergent magneto-electronic sensors has been a result of decades-long research on physical and chemical processes that underpin a reversible absorption of hydrogen by Pd [18]. Indeed, when a hydrogen molecule approaches the surface of Pd, the H-H interdistance increases because of a strong Pd-H interaction, thereby weakening the H-H bond. Then, the hydrogen molecule dissociates into two unbound hydrogen atoms. Since the atomic radius of hydrogen is small (approximately 0.53 Å), hydrogen atoms can easily diffuse into Pd. Initially, hydrogen atoms occupy the interstitial sites of the crystalline lattice (α-phase). However, as hydrogen absorption increases, hydrogen atoms expand the crystalline lattice (β-phase). Usually, hydrogen atoms in the α-phase are more stable and harder to desorb than those in the β-phase.

It has also been long known that during absorption of hydrogen certain physical properties of Pd change and that these changes can be used to detect the presence of hydrogen and its concentration. In particular, as a result of hydrogenation, the atomic lattice parameter of Pd can increase by up to 3% and the electrical resistivity of Pd by up to 80%. Pd was also found to be highly selective to $H_2$ absorption, but at the same time, it exhibits a much lower sensitivity to other gases [10], though they can still poison a Pd-based sensor and therefore special protection measures may need to be taken [8]. This prior knowledge has significantly influenced the course of development of the emergent magneto-electronic hydrogen gas sensors.

*1.4. Motivation to develop alloy-based magneto-electronic hydrogen gas sensors*

Whereas Pd has been employed in many hydrogen gas sensor architectures, a combination of hydrogen-induced properties of Pd with the typical material, structures and experimental techniques used in magnonics, spintronics, magneto-electronics and adjacent areas (Section 2) is a result of very recent developments. Naturally, to optimise the operation of the first prototypes of magneto-electronic sensors and to further demonstrate their plausibility and applicability in a wide range of practical situations, the researchers working on them could refer to the works on the previous generation of Pd-based hydrogen sensors. However, as for example stated in [17], the analysis of the available literature sources revealed that "*whether new magnetic systems highly sensitive to the hydrogen absorption can be developed and whether such system can be applied to hydrogen sensing or magnetism control remained unexplored and thus warranted investigation*". Consequently, pioneering research efforts had to be done to produce new fundamental and technical knowledge in this area.

Among the previous works that have facilitated the development of the emergent magneto-electronic sensor technology were, for example, the papers on PdNi alloy thin-film electric-resistivity based gas senors [19, 20] intended to improve the performance of commercial sensing technologies in several important ways. In particular, those works suggested that the use of alloys could help decrease the size of the device and optimise its production cost due to a potential of using batch fabrication methods. Yet, a compatibility of the techniques of fabrication of alloy-based sensors with the standard silicon device manufacturing processes was expected to help achieve sensor temperature control and signal processing, including networking of many sensors, on a single integrated chip, thereby leading to a greater system reliability and to an even lower cost. The solid-state format with a built-in chip heater was also projected to enable $H_2$ sensing over a very large ambient temperature range, including leaks of cryogenic $H_2$ used in rocket motors. Finally, such sensors were estimated to reach a sub-second response time.

To create such alloy-based sensors, research efforts were concentrated on measurements of higher concentrations of $H_2$ and on alloy formulation that would enable a reversible detection of the gas. The role of temperature relative humidity and of oxygen on the performance of the sensor alloys were also studied in great detail. For example, Fig. 1(a) shows a hydrogen-induced response of a 50-nm-thick film of a PdNi alloy, where a fast and reversible percent change in the resistivity due to changes in the concentration of $H_2$ can be observed. Figure 1(b) shows the results of a test of the effect of sensor poisoning by $H_2S$, where one can see an insignificant effect of poisoning on the response rate of the device.

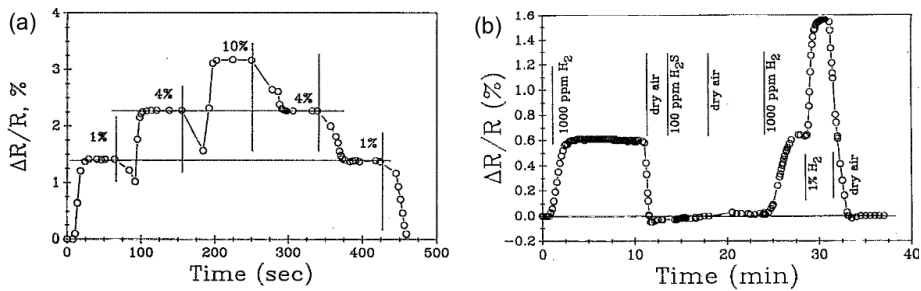

**FIG. 1**: **(a)** The response of a 50-nm-thick film of PdNi alloy (8% of Ni) demonstrating a fast and reversible relative change in the resistivity due to changes in the concentration of $H_2$. The vertical lines indicate the time corresponding to a resetting of the flow controllers for the new gas mixture. After the final measurement, the sample was purged with dry air. **(b)** Effect of poisoning by $H_2S$ on the PdNi alloy sensor. Reproduced with permission from [19].

In another work Ref. [21], PdNi alloy nanoparticles with a controlled Ni atom content ranging from 0 to 60%, were synthesised, where the mean size of the alloy nanoparticles was found to decrease with an increase in the Ni content. Then, electric conductance-based hydrogen sensors were fabricated by depositing films of closely spaced PdNi nanoparticles. An optimal Ni content of 16% was suggested to create $H_2$ sensors with a rapid response, high sensitivity as well as good linearity and reliability in a wide hydrogen pressure range. Note that Ni is a ferromagnetic metal and, whereas its ferromagnetic properties may not be directly relevant to the concept of hydrogen gas sensing described above, it is known that alloys of Pd with an FM exhibit significant magnetic properties that can be exploited in magneto-electronic gas sensors.

Thus, inspired by the results of the aforementioned works on the alloys and also by developments in the adjacent areas, it has been suggested that the hydrogen-sensitive thin films could be made of an alloy of Co or Fe with Pd [17, 22]. In fact, changes to the magnetic state of the alloy induced by the presence of $H_2$ gas in the film environment could be used to detect the gas presence, and understanding of this existed earlier because of the well-established fact that an alloy could be formed at an interface between Pd and FM layers [23], as explained in [24]. Most likely, similar concepts had been also put forward by the researchers working on magneto-electronic hydrogen sensing exploiting a magneto-optical [12] and a Hall-family effect-based mechanism [25] of the sensor state registration.

However, whereas Pd-FM interfaces have also been central for thin-film structures used in the fields of spintronic, magnonics and data storage, and therefore considerable understanding of the physics underpinning their operation has already been gained (for an extended relevant discussion see, e.g., [16]), at the time of the first proposals of magneto-electronic sensors the perspectives of employing alloys as a sensitive material were still not well-established. That situation motivated further studies of magneto-electronic properties of alloys of Pd with an FM metal and of the effect of hydrogenation on them. However, the outcomes of those detailed studies have been scattered in the literature, and hence they are often difficult to access by students and investigators of hydrogen gas sensors based on other concepts than magneto-electronics. Thus, the current work is an attempt to systematically explain the physics of operation of magneto-electronic sensors using alloys and to critically overview the main results achieved in this field of research to date.

We note the existence of a recent review article [16], where the interested reader will find detailed information about the general physical and engineering aspects of magneto-electronic sensors. However, although the paper [16] also presents some aspects of alloy-based magneto-electronic sensors, it is written mostly with multilayered (i.e. non-alloyed) structures in mind. Subsequently, the discussion provided hereafter not only complements the previous paper [16] but rather aims at introducing the concept of magneto-electronic sensors to a wider research community interested in alloyed materials and their applications in the field of gas sensing. In line with such a goal, we organise the following discussions so that they could be read independently of any other work on magneto-electronic sensors and relevant concepts at least at the introductory review level. At the same time, we believe that the text would also be of use to those seeking specific information. To that end, our discussion is supported by a unique selection of cited literature sources.

## 2. Physical processes underpinning the operation of magneto-electronic gas sensors

Here, we overview the fundamental physical processes that underpin the operation of magneto-electronic devices, and we also discuss some of the key results presented in the seminal works on magneto-electronic hydrogen gas sensors using the effect of Ferromagnetic Resonance, magneto-optical Kerr effect and relevant physical phenomena. Whereas mostly multilayered thin films were investigated in those works, the results obtained using them are of immediate relevance to the physics of operation of alloy-based magneto-electronic sensors. Therefore, they can be considered to be a starting point.

*2.1. FMR-based magneto-electronic hydrogen sensors*

Ferromagnetic Resonance (FMR) is an effect of spin precession of the macroscopic vector of magnetisation in an external magnetic field [26]. The eigen-frequency of precession lies in the microwave range (0.5 to 100 GHz). Therefore, in an FMR experiment, the spin precession is usually driven by an external source of microwave power [26], and the onset of the resonance is observed as an increase in absorption of microwave power by the sample, when the frequency of the external source matches the FMR frequency.

The concept of hydrogen gas sensor exploiting FMR was originally proposed in [13], where a change in the FMR peak position upon absorption of $H_2$ by a Pd/FM-metal bilayer thin film was measured using a broadband stripline FMR spectroscopy technique [26]. The experimental setup used in [13] consisted of a customised air-tight cell (Fig. 2(a), for more technical details see [27]) that enabled controlling the flow of gas at atmospheric pressure along a co-planar stripline waveguide (CPW) and the sample under study placed above it. The coaxial cables were used to supply microwave power from a generator to the CPW input port and to carry the power from the CPW output port towards a microwave receiver. The cell was fixed between the poles of an electromagnet so that the magnetic field was applied along the CPW, which is the orientation that enables maximising the FMR response. Figure 2(b) shows representative FMR spectra measured for a Co(5nm)/Pd(10nm) film under $N_2$ and $H_2$ atmospheres and it demonstrates a readily detectable shift of the FMR peak towards the lower applied magnetic fields upon absorption of $H_2$.

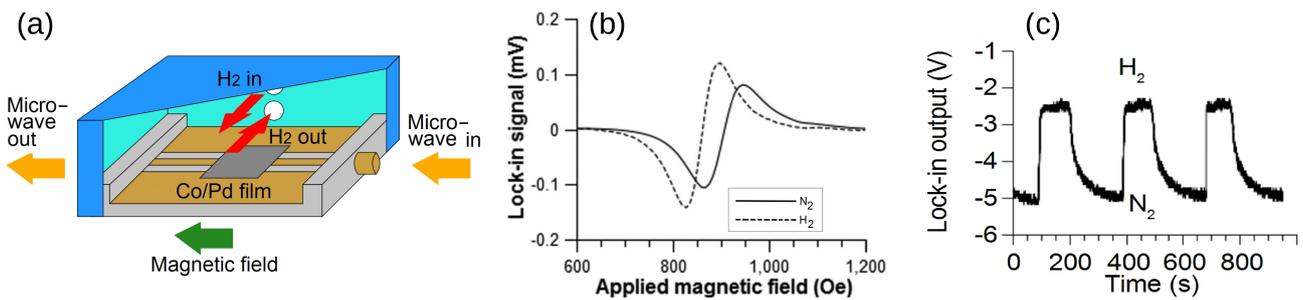

**FIG. 2**: **(a)** Sketch of the gas cell used to demonstrate the plausibility of the concept of FMR-based magneto-electronic sensor. The sketch shows the co-planar stripline waveguide (CPW), where the sample is located, the microwave feed ports and the gas flow inlets. **(b)** The FMR spectra of the Co/Pd film at 10 GHz microwave frequency in $N_2$ atmosphere (solid line) and $H_2$ atmosphere (dashed line). **(c)** Changes in the CPW output voltage under the cycling of $N_2$ and $H_2$ gas through the cell with the Co/Pd film operating at the FMR frequency. Reproduced with permission from [13].

The FMR-based approach illustrated in Fig. 2(a) has a number of advantages over other methods of registration of the state of a magneto-electronic sensor that we discuss below. Firstly, the fabrication of magnetic multilayered films involving metals of Platinum group (Pt, Pd, etc.) is a well-established and inexpensive procedure that employs the techniques used to manufacture magnetic hard drives for computers. Secondly, the FMR spectroscopy is a technically simple and convenient tool for reading the state of the sensors. Indeed, as can be seen in Fig. 2(b), not one but two aspects of the sensor state simultaneously change in the presence of $H_2$ gas – the position of the FMR peak and the width of the FMR resonance line. The FMR linewidth is a measure of magnetic losses in a magnetic material. Thus, a hydrogen-induced decrease in the resonance linewidth implies that absorption of hydrogen can reduce the rate of energy loss by a particular magnetisation precession mode. (Generally, the precession energy is lost due to dynamic processes that couple this particular magnetisation precession mode to other degrees of freedom of the spin system in the material.) Importantly, a decrease in the linewidth, in turn, must yield an increase in the amplitude of the FMR peak, which is an effect based on laws of electrodynamics. The fact that the amplitude of the FMR peak increases in the presence of $H_2$ gas in the sample environment is clearly seen from the Fig. 2(b). Significantly, both the FMR peak shift and the changes to the FMR peak width and amplitude are fully reversible and repeatable [Fig. 2(c)], which is an essential property for any hydrogen gas sensing application.

In Ref. [13], it was also demonstrated that the sensor state can be registered remotely by means of FMR measurements conducted through an optically non-transparent and electrically-insulating wall of a gas cell containing $H_2$. This is in contrast to the magneto-optical techniques (Section 2.2) that have their own certain advantages in the context of magneto-electronic hydrogen gas sensing, but require a specially designed and fabricated optically transparent window, the presence of which may complicate the fabrication process and increase the cost of the device. Finally, in contrast to many Pd-based optical [28], electrical [29] and micro-mechanic [30] sensors, the FMR-based magneto-electronic sensor approach does not require mechanical stretching/shrinking of the Pd layer (for a more detailed discussion see [16]). It is well-known that the presence of a hydrogen-induced strain decreases the lifetime of the sensor due to irreversible layer deformations [31, 32] and hysteretic sensitivity [32]. Strain-based sensing also requires large Pd thicknesses to overcome the substrate clamping effect [32], thereby increasing the cost of the sensor, since industry-grade Pd is relatively expensive. On the contrary, the magnetic hydrogen gas sensing is not necesserily based on mechanical degrees of freedom that relaxes requirements to the film thickness. Below we will return to this point.

The FMR peak position for an ellipsoid of revolution is given by Kittel formula [33]:

$$f^2 = [\gamma/(2\pi)]^2[H + (N_{xx} - N_{zz})M_s][H + (N_{yy} - N_{zz})M_s], \qquad (1)$$

where $f$ is the FMR frequency in Hz, $\gamma$ is the gyromagnetic ratio, $H$ is the static magnetic field applied to the sample, $M_s$ is the saturation magnetisation for the sample and $N_{xx}$, $N_{yy}$ and $N_{zz}$ are the demagnetising factors for the ellipsoid. Equation (1) assumes that the static field $H$ is applied along the $z$-direction and that the sample is magnetised uniformly (i.e. to saturation). Importantly, it must stand $N_{xx} + N_{yy} + N_{zz} = 4\pi$.

A thin film is a special case of the ellipsoid of revolution, for which one of the demagnetising factors equals to $4\pi$ and the other two vanish. Then, for two special cases of the orientation of the static magnetic field with respect to the film plane, Eq. (1) reduces to

$$f^2 = [\gamma/(2\pi)]^2 H(H + 4\pi M_s - H_{PMA}) \qquad (2)$$

for the static magnetic field applied in the film plane ("in-plane" or IP FMR) and,

$$f = [\gamma/(2\pi)](H - 4\pi M_s + H_{PMA}) \qquad (3)$$

for the static magnetic field applied perpendicular to the film plane ("perpendicular-to-plane" (PP) or "out-of-plane" (OOP) FMR). While writing down Eq. (2) and Eq. (3), we included the effective field of perpendicular magnetic anisotropy $H_{PMA}$ into the expressions. (However, we did not include $H_{PMA}$ into Eq. (1) to keep that equation simple enough.)

As seen from Eqs. (1-3), the FMR peak position depends on the applied static magnetic field $H$, the value of the saturation magnetisation of the constituent material of the film $M_s$ and of the effective field of PMA $H_{PMA}$. The latter quantity requires explanation, and we will provide it later on in several contexts.

Now we focus on Eq. (2) and Eq. (3), from which one can see that, in an experiment, the position of the FMR peak may be found following two different approaches. The first one is by keeping the applied magnetic field $H$ constant and sweeping the frequency $f$ of a microwave generator that drives FMR across the FMR resonance frequency. Alternatively, one can keep $f$ constant and sweep

*H*. The latter approach is technically less challenging in general and therefore it is used more often. For instance, the FMR traces in Fig. 2(b) were obtained using this method, with microwave frequency kept at *f* = 10 GHz. In addition, one more technical approach – the field-modulated FMR – was utilised in combination with the field sweeps to record the traces shown in Fig. 2(b). Its goal was to further increase the sensitivity of the experimental setup (for specific technical details see Ref. [26]). A side effect of this method is that instead of registering a peak with a Lorentzian shape that is a typical resonance peak shape for an externally driven resonance, in the case of a field-modulated FMR the registered peak takes the form of the first derivative of the Lorentzian shape.

One more aspect needs to be explained before we are able to proceed further. This is the choice of the FMR configuration – IP or OOP. As seen from Eq. (2) for the IP FMR, the dependence of the FMR frequency on the applied static magnetic field *H* is not linear. As a result, extraction of the magnetic parameters such as the film saturation magnetisation $4\pi M_s$, the gyromagnetic ratio γ and the PMA field $H_{PMA}$ from the raw FMR traces is less unambiguous [34]. Conversely, in the OOP configuration, the dependence is linear with a slope given by the value of *γ* and with the *x*-axis intercept by $4\pi M_s - H_{PMA}$. The latter quantity is often called the effective magnetisation $4\pi M_{eff}$. A drawback of the OOP configuration for the $H_2$ gas sensing is that much larger static magnetic fields have to be applied for the same FMR frequency than in the case of IP FMR. On the other hand, the OOP FMR ensures a much larger sensitivity to a change in $4\pi M_{eff}$ as shown, for example, in Ref. [35]. Finally, from both Eq. (2) and Eq. (3), one can see that the FMR experiment is sensitive to $4\pi M_{eff}$ but it is unable to separate the contributions of $4\pi M_s$ and $H_{PMA}$ to the FMR peak position. This implies that while studying the effect of hydrogen gas on the FMR response of an FM film, one cannot infer whether $4\pi M_s$ or $H_{PMA}$ or both have been affected by hydrogen absorption, unless one possesses some prior knowledge of it.

Now we return to the discussion of PMA, which is a type of magnetic anisotropy, and clarify the physical origin of this effect. By definition, the magnetic anisotropy is a dependence of magnetic energy of an FM material on the direction of its magnetisation vector with respect to particular directions in the sample. In the main discussion in this text, we are dealing with polycrystalline metal films of very large in-plane dimensions with respect to the film thicknesses. Subsequently, there is just one special direction in this geometry – the one perpendicular to the film plane. The dependence of the film magnetic energy on the angle of the magnetisation vector with the film normal is called the Perpendicular Magnetic Anisotropy or PMA[1].

Furthermore, there are two types of PMA in FM films – the interface PMA and the bulk PMA. The interface PMA exists at the interfaces of FM metals with metals of Pt group, including Pd [36, 37, 38]. This type of PMA can significantly affect the FMR response of a bilayer or multi-layer FM-metal/Pd film only if the thickness of the FM layer is small, say, less than 10 nm or so. The thinner the layer the stronger the impact of the interfacial PMA on the FMR response. The effect of hydrogen on the interface PMA was exploited in the multilayer-based magnetic hydrogen sensing in Refs. [13, 14, 35, 39, 40]. An interested reader can find further details of the effect of $H_2$ gas on the interfaces of Pd with FM metals in Refs. [24, 41, 42, 43, 44, 45, 46, 47, 48, 49].

On the contrary, the bulk PMA exists in the bulk of an FM layer and it originates from a break in symmetry of the spin system induced by growth of the film on its substrate. A particular orientation of film crystallites or a strain induced in the film during the growth are among the origins of this type of anisotropy. Importantly, the effect of the bulk anisotropy on the FMR response does not depend on the film thickness. One example of materials where bulk PMA may be present are the films of alloys of FM metals with Pd. Note that ultimately, on the microscopic level, any magnetocrystaline anisotropy is due to spin-orbit coupling, and that macroscopic effects of the preferential crystallite orientation or internal strain are able to modify the spin-orbit coupling such that macroscopic PMA is induced.

---

[1] It is important not to confuse PMA with the shape anisotropy of the film. The shape anisotropy is a pure magnetostatic effect that is fully described by the demagnetising factors for the film [see Eq. (1)]. Conversely, PMA originates from processes in the material on the mesoscopic and microscopic levels and represents a contribution to the total magnetic energy *on top* of the trivial contribution of the shape anisotropy energy.

*2.2. Magneto-optical Kerr effect-based magneto-electronic hydrogen sensors*

The FMR spectroscopy is not a unique way to register the state of a magneto-electronic sensor. In fact, magneto-electronic hydrogen sensors using the magneto-optical Kerr effect (MOKE) have also been developed alongside the FMR-based sensors (see, e.g., Refs. [12, 50, 51, 52, 53]) and the knowledge obtained using MOKE techniques has benefited and complemented the one obtained using FMR.

MOKE represents a change in the polarisation and intensity of light reflected from the surface of a magnetic material. The change originates from the off-diagonal components of the dielectric permittivity tensor of the investigated magnetised material [54]. Depending on the direction of the sample magnetisation with respect to its surface and the plane of incidence of light, MOKE can be observed in polar, longitudinal and transverse configurations [54, 55]. All three MOKE configurations can be used to characterise magnetic materials. Given that MOKE can be observed mostly in optically reflecting materials, MOKE-based magnetometry techniques have been especially suitable for studying magnetism of strongly optically-reflecting metals [56, 57, 58, 53].

In one of the works that laid a foundation of the MOKE-based magneto-electronic hydrogen sensors, a reversible nature of the changes in the magnetic properties was first observed in Pd/Fe bilayer structures placed into an $H_2$–containing atmosphere [12]. In that experiment, the thickness of the Pd layer was increased and the MOKE setup gradually adjusted until a point, where a considerable enhancement of the intensity of the MOKE signal was reached after the exposure to 1 atm of $H_2$. The reversibility of this change was confirmed by cyclic desorption and reabsorption of hydrogen, thereby revealing a practically important sensitivity of the magneto-optical response of a material combining an FM metal (Fe) with a highly hydrogenated NM metal (Pd). Similar results were obtained in the follow-up works [50, 44] on Pd-based ferromagnetic multilayers. A reader interested in a continuation of this particular discussion is referred to a review article [16].

*2.3. Anomalous Hall effect*

The term anomalous Hall effect (AHE) [59, 25, 60, 61] originates from the ordinary Hall effect [Hur72]. In an FM material, there is an additional contribution to the Hall voltage known as the anomalous or extraordinary Hall effect [59, 25, 60, 61, 62, 63]. This voltage difference contribution depends on the value of material magnetisation and in some materials it can be larger than the ordinary Hall voltage [59, 60]. Since the strength of the anomalous Hall effect is a function of spin-dependent scattering of the charge carriers, a change in the magnetic state and electrical conductivity of the sample results in a change in the total Hall voltage [59, 25, 60, 61, 63].

These physical processes lay a foundation of a special kind of magneto-electronic hydrogen gas sensors that are intrinsically compatible with existing electronic gas detection technologies and that enable a simultaneous measurement of two independent parameters affected by $H_2$ – resistivity and magnetisation. It has also been demonstrated that using the anomalous Hall effect one can create a device capable of detecting two gases – $H_2$ and CO [64]. A more detailed discussion of those results is given in Section 4.

*2.4. Impact of hydrogenation on magnetic properties of Pd-based FM alloy films*

Before we start discussing applications of Pd-based alloyed films in magneto-electronic hydrogen gas sensing, it is worth making the reader familiar with the physics underlying the impact of $H_2$ gas on the magnetic properties of alloyed thin films. Many of the material-science-orientated works discussed in this section were published after the introduction of the concept of magneto-electronic hydrogen gas sensing based on the alloyed films. However, alongside the predated works, the results presented in them help explain the available results on magneto-electronic $H_2$ sensing and to guide the future research in this field.

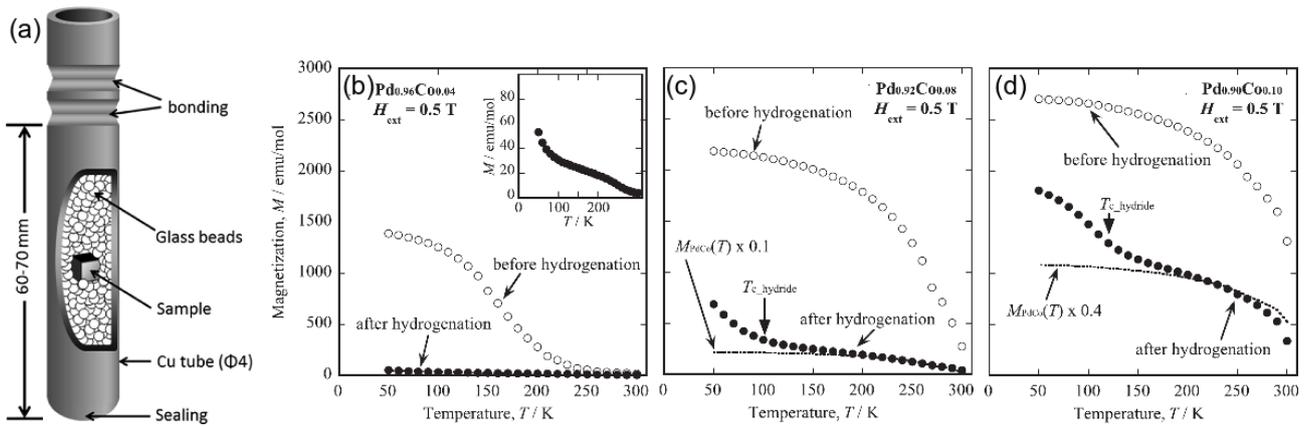

**FIG. 3**: **(a)** Sketch of a sample holder in the SQUID (superconducting quantum interference device) magnetometer used in the experiments. **(b-d)** Temperature dependence of the magnetisation for each investigated sample obtained at the external magnetic field of 0.5 T. The open and closed circles denote the magnetisation-vs-temperature curves taken before the hydrogenation and after it, respectively. The inset in panel (a) shows the data with an enlarged *y*-axis of the hydrogenated $Pd_{0.96}Co_{0.04}$ sample. Reproduced with permission from [65].

It is well-known that alloys of Pd with a FM metal exhibit intriguing and practically important magnetic properties [23] that can also be relevant to magneto-electronic hydrogen gas sensing. For example, the magnetic properties of PdCo alloys were investigated in [65], where temperature dependencies of the magnetisation of PdCo alloy samples were used to observe a ferromagnetic behaviour and to estimate Curie temperature (Fig. 3). Then, the PdCo alloys were hydrogenated under 100 kPa $H_2$ gas pressure and the resulting temperature variations of the magnetisation revealed two magnetic transitions implying that there were two phases in the hydrogenated PdCo– the hydrogen dissolution and hydride phases.

Hydrogen-induced changes in the interaction between magnetic layers were also investigated in Fe/Nb and Fe/V multilayer structures and they revealed a transition from an antiferromagnetic to a ferromagnetic coupling regime [66, 67, 68]. In a number of later works, $Fe_xPd_{1-x}$ and $Co_xPd_{1-x}$ alloys were investigated for their reversible magnetic properties caused by the hydrogenation [69, 70, 71, 65, 73, 17]. To further investigate the fact that hydrogenation can be used to modulate the magnetic properties of Pd-rich FM alloy films, Pd-rich Mn/MnPd/Fe antiferromagnetic/FM films were studied in Ref. [73], where it was shown that hydrogenation can indeed enhance the exchange bias coupling between the layers. Specifically, magnetic hysteresis loops revealed that the magnetic state of the Pd-rich Mn/MnPd/Fe films could be gradually changed from an unbiased state to an exchange-biased state upon exposure to $H_2$ gas. Yet, hydrogenation promoted a larger magnitude of the exchange bias field, which was attributed to an enhanced long-range antiferromagnetic ordering of the Pd-rich MnPd alloy film caused by the uptake of hydrogen atoms. Relevant studies were also conducted in the context of optical surface plasmon polariton (SPP)-based hydrogen sensors [74].

Furthermore, it is well-known that at high $H_2$ concentrations, the Pd atomic lattice experiences a significant expansion [75, 76]. When Pd serves as a constituent material of a thin film clamped to a substrate, its lattice cannot expand horizontally but only vertically. Thereby large internal stresses are induced in the film. Whereas Pd layers with thicknesses on the order of 10 nm can easily withstand such stresses, layers with thicknesses in the range above 30 nm can develop cracks and other signs of mechanical deterioration [32]. Yet, as discussed, for example, in Ref. [16], elastic stresses and the magneto-elastic PMA effects as well as accompanying effects such as magnetostriction [77] play an important role in the operation of FMR-based magneto-electronic sensors. Similar effects have also been investigated in PdCo and PdNi alloys in the presence of $H_2$ gas [51, 78, 79, 80, 81] and, therefore, they are of immediate relevance to magneto-electronic hydrogen sensing. Hence, we discuss those results in more detail.

There exist two major physical mechanisms of the impact of hydrogen on the magnetic properties of PdCo and FeCo alloy films. The first one is a modification of elastic stress in the films upon absorption of hydrogen [79, 80]. In Ref. [80] it was found that the pristine-state $Pd_{0.85}Co_{0.15}$ films are characterised by a magnetic uniaxial anisotropy. Those films had a (111) structure that is typical for fcc metal films deposited on glass substrates, where the anisotropy originates from an elastic stress induced during the sample growth on its substrate and clamping of the film to the substrate. In the pristine samples the stress was compressive; it induced uniaxial anisotropy through the magnetostriction effect. That anisotropy was of easy-plane type (it was called longitudinal magnetic anisotropy (LMA) in [80]).

Following the authors of [80], annealing the alloy would result in a shrinkage of its crystal lattice if the material were not clamped to the substrate ("free state"). However, because the film is clamped to the substrate ("clamped state"), significant in-plane shrinking is impossible. The crystal unit cell then appears to be stretched with respect to its free-state size. The film is able to remain stretched due to a tensile stress induced by the clamping effect. As a result, the inability of the clamped film to shrink converts the original compressive stress into a tensile one. Accordingly, LMA converts into a magnetostriction-induced PMA.

As already mentioned, absorption of $H_2$ gas is also known to lead to an expansion of the crystal lattice of pure Pd metal and its alloys. Films clamped to substrates cannot expand in the film plane, but they are able to expand in the vertical direction, which was confirmed by X-ray diffraction (XRD) in Ref. [80]. As a result of these processes, a compressive component is added to the original stress in the films. For the as-deposited films, this increases the already present compressive stress leading to a stronger LMA upon loading the films with hydrogen. Similarly, for the annealed films, adding a compressive component reduces the original tensile stress leading to a decrease in PMA. After desorption of hydrogen, all investigated films recovered their original anisotropies.

The authors of [80] also conducted Density-Functional Theory (DFT) simulations that showed that Pd atoms are magnetically polarised due to their proximity to Co ones. Hydrogenation of the alloy results in a small decrease of the magnetic moment of Pd. This decrease does not affect significantly the saturation magnetisation value of the alloy, but it can noticeably decrease magnetostriction, because polarised Pd atoms play an important role in magneto-elastic coupling.

The authors of another relevant work Ref. [51] investigated the magnetic properties of a multilayer film representing $Co_{40}Pd_{60}$ layers separated by Cu spacers. They used an X-ray absorption near edge structure (XANES) spectroscopy technique to probe magnetic moments at Co and Pd atoms. They established that the incorporation of hydrogen into the crystal lattice of CoPd results in a hydrogen-mediated charge transfer from Co to Pd via *p-d* orbital hybridisation. The charge transfer is a spin-dependent process because it removes electrons from the minority band of Co. Subsequently, it results in a decrease in the saturation magnetisation of the alloy. In addition, they saw evidence in support of a distortion of the PdCo unit cell upon absorption of hydrogen that is consistent with the expected expansion of the crystal lattice. Thus, it was concluded that magnetism of the alloy can also be altered through alternation to Co-Pd and Co-Co bonds.

Finally, in Ref. [78] an X-ray Magnetic Dichroism (XMCD) technique was employed to investigate the effect of hydrogen on the electronic structure of FePd alloys. It was established that the magnetic moment of Fe in a $Fe_{40}Pd_{60}$ alloy film is increased upon absorption of $H_2$ gas. An increase in the ratio of orbital to spin moment at Fe atoms was also found, and it was argued that the presence of hydrogen does not change the spin moment at Fe but increases its orbital moment potentially due to a distortion of the crystal lattice.

## 3. Application of alloys in MOKE-based magneto-electronic hydrogen gas sensors

Alloys of Pd with Co, Fe and Ni have been intensively investigated for applications in MOKE-based hydrogen gas sensing [17, 82, 52, 78]. Apart from the motivation to investigate alloys in the context of hydrogen sensing already discussed in Section 1.4, the use of MOKE has also been intended to resolve some of the fundamental and technological challenges faced by the investigators of FMR-based alloy magneto-electronic hydrogen sensors. In particular, while using Pd alloys in FMR-based hydrogen detection, one observes an increase in the linewidth of the FMR peak and a concomitant decrease in the amplitude of the peak [13], which is a well-known fact established in FMR measurements of samples made of impure FM metals [26]. This disadvantage of the FMR-based detection approach was explicitly mentioned in [17] as one of the key reasons for conducting their MOKE-based experiments, the key results of which we overview in continuation.

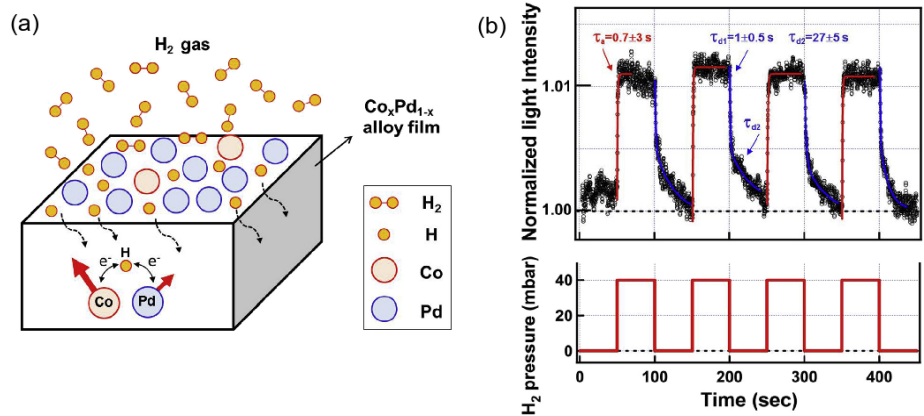

**FIG. 4**: **(a)** Sketch of the process of hydrogen gas dissociation on the surface of a $Co_xPd_{1-x}$ alloy film and hydrogen gas absorption and diffusion into it. The hydrogen atoms may cause charge transfer thereby resulting in substantial changes in the magnetic properties of the film. **(b)** Observations of reversible variations in the intensity of light reflected from a 15.5-nm-thick $Co_{24}Pd_{76}$ alloy film measured using MOKE under a cyclic exposure to 40 mbar $H_2$ gas (the bottom panel). The open circles denote the experimental data points but the solid curves represent the fitting results used to extract the time constants of the hydrogen absorption ($\tau_a$) and desorption effects ($\tau_{d1}$ and $\tau_{d2}$). Reproduced with permission from [17].

Earlier studies of Pd/Co/Pd trilayers by the same research group as in the paper Ref. [17] revealed a noticeable and reversible change in the magnetic coercivity induced by hydrogenation. An annealing-induced CoPd alloy interface was suggested to be the cause of that hydrogenation effect [Fig. 4(a)]. To further investigate that effect, in the work [17] a batch of CoPd alloy thin films with varying alloy compositions were studied. MOKE hysteresis loop measurements of the samples with a dominating Co content revealed the presence of an in-plane anisotropy. Subsequently, exposure of those samples to $H_2$ gas did not produce observable changes to the shape or intensity of their hysteresis loops. On the contrary, samples with a dominant Pd content displayed PMA that, in turn, enabled observations of significant changes in the MOKE hysteresis loops when the samples were exposed to $H_2$ gas. Those results are in good agreement with our earlier discussions of the key role of PMA in magneto-electronic hydrogen gas sensors (for example, see Section 2.1).

It is noteworthy that in the MOKE measurements discussed above, a *p*-polarised 670 nm laser beam was focused on the sample and an analyser, consisting of a linear polariser oriented at a small angle φ with respect to the *s*-axis and of a photodiode, was used to detect the reflected light [83]. Because of the general magneto-optical properties [54, 55, 83] of the CoPd alloy films, the reflected beam was not purely *p*-polarised but comprised both a *p*-component and an *s*-component. The ratio of the optical intensities of these components and the knowledge of the angle φ enables one to extract the Kerr rotation and the Kerr ellipticity from the experimental data. In a typical experiment, the light intensity $I_{\pm m}$ was also observable as a function of the angle φ for a positively (+*m*) and a negatively (−*m*) magnetised sample. As a result of hydrogenation, both the light intensity and optical extinction angle changed. This allows observing effects that can be used to reliably detect hydrogen gas in a wide range of concentrations.

The magnetic properties of the alloys films also depend on the film thickness. To demonstrate such a dependence, MOKE hysteresis loops were measured for $Co_{39}Pd_{61}$ alloy films with the thicknesses of 10, 14.8 and 20 nm as a function of hydrogen pressure (Fig. 5). It was found that for the 10-nm-thick alloy film, the magnetic anisotropy tended to be an in-plane one for both pristine and hydrogentated states. Conversely, for the thicker films, because of the volume-contributed perpendicular anisotropy, the pristine magnetic easy axis of the thin films transferred from the in-plane to the perpendicular direction as the film thickness increased. Loading the films with hydrogen yielded enhancement of coercive field for the films, and the thicker the film, the larger the enhancement. Importantly, an increase in squareness of the MOKE hysteresis loops was also observed. The squareness increased in both perpendicular and in-plane magnetisation directions. Therefore, possible reorientation of the easy axis from IP to OOP could be excluded. This prompted the authors to conclude that a transition from a disordered and short-range magnetic coupling state to a long range-ordered ferromagnetic state takes place upon loading the films with hydrogen and that hydrogen atoms in the interstitial space between Co and Pd atoms are critical in effecting this transition.

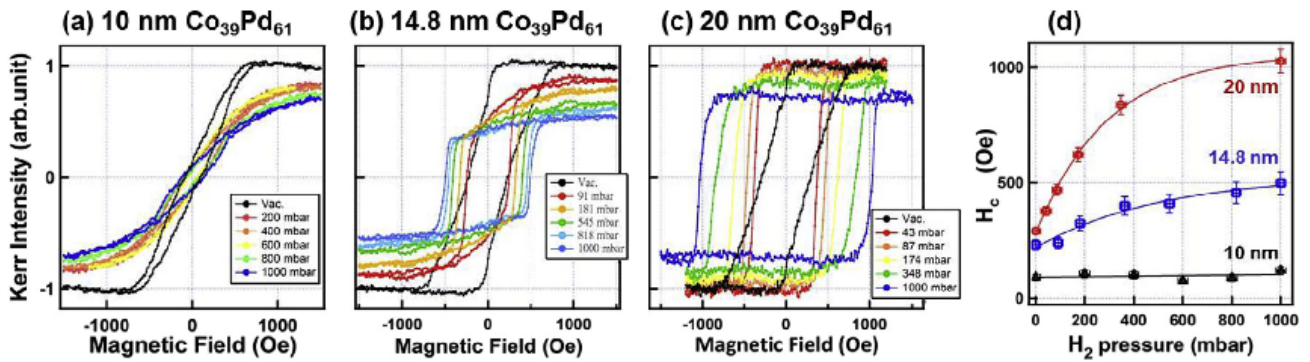

**FIG. 5**: Effect of hydrogenation on the perpendicular MOKE hysteresis loops of $Co_{39}Pd_{61}$ alloy films of the thickness (a) 10 nm, (b) 14.8 nm and (c) 20 nm. In panel (d), the perpendicular magnetic coercivity of these films is plotted as a function of $H_2$ gas pressure. Reproduced with permission from [17].

Figure 4(b) demonstrates the reversibility and time-dependence of the hydrogenation effect on a 15.5-nm-thick $Co_{24}Pd_{76}$ alloy film from the same Ref. [17]. The MOKE measurements were conducted in a test gas cell that was cyclically pumped to a vacuum and filled with $H_2$ gas [the bottom panel of Fig. 4(b)]. After the exposure to 40 mbar of $H_2$, the intensity of light reflected from the sample immediately increased to the saturated maximum, but after the pumping the $H_2$ gas out of gas cell the light intensity first reduced rapidly in a few seconds and then it continued to decrease a relatively slow pace for a longer period of time.

A curve fitting procedure was employed to extract the time constants of hydrogen absorption ($\tau_a$) and desorption ($\tau_{d1}$ and $\tau_{d2}$) from the data plotted in Fig. 4(b). For the absorption stage $\tau_a = 0.7 \pm 0.3$ s was obtained. For hydrogen desorption, a double exponential function had to be used to correctly fit the data, resulting in the rapid reduction time $\tau_{d1} = 1 \pm 0.5$ s and the slower reduction time $\tau_{d2} = 27 \pm 5$ s. The presence of the two time constants was attributed to a difference in hydrogen desorption rates for the β- and α-phase. This is qualitatively consistent with the results reported in Ref. [71], where it has been demonstrated that alloy thin films responded much faster than pure Pd films or Pd-based multilayers. The fact that the response time observed in the CoPd alloy films is shorter than that in NiPd alloy films was attributed to variations in surface morphology and crystalline structure of the constituent materials [71].

Continuing the discussion of the application of alloys of Pd with FM metals, it should be noted that whereas the effect of sensitivity to hydrogen has been demonstrated to have a high application potential, the correlation between the hydrogen pressure, ambient temperature and magnetic properties of the alloy still remains insufficiently understood from the physical point of view. To address this challenge, in a recent work Ref. [78] the magnetic moment of Fe in an FePd alloy thin film was found to increase through absorption of hydrogen as evidenced by observations of an enhanced X-ray magnetic circular dichroism signal of Fe. Hydrogen absorption and desorption MOKE hysteresis loops [Fig. 6(a)] were also analysed revealing a reversible character of the absorption-desorption process [Fig. 6(b)]. It was also suggested that at temperatures around 300-360 K the hydrogen bonding to the alloy film could be modulated using thermal activation without significant hydrogen desorption, thereby providing an insight into the role of temperature in the effects of hydrogen on magnetism [Fig. 6(d-e)]. These results should be valuable for future applications in magneto-electronic hydrogen gas sensors and concentration meters.

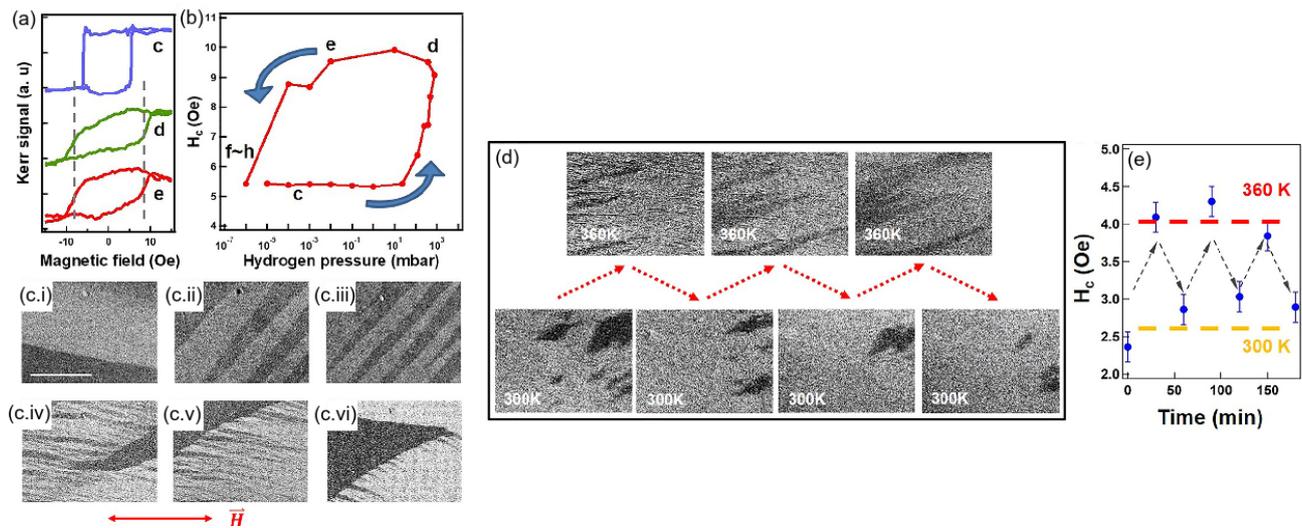

**FIG. 6**: Left: Hydrogen-induced variation of the magnetic coercivity $H_c$ in a 55-nm-thick $Fe_{40}Pd_{60}$ alloy film. **(a)** MOKE hysteresis loops corresponding to the data points $c$, $d$ and $e$ in panel (b). **(b)** Hysteretic behaviour of $H_c$ as a function of $H_2$ gas concentration. The panels **(c.i)–(c.vi)** are the magnetic domain profiles corresponding to the data points from $c$ to $h$ in panel (b). **Right: (d)** Results obtained after hydrogenation under 1 bar $H_2$ with a cyclic temperature switching between 300 and 360 K. Reproduced with permission from [78].

### 4. Alloy-based magneto-electronic sensors exploiting the anomalous Hall effect

The term anomalous Hall effect (AHE) [59, 25, 60, 61] originates from the ordinary Hall effect [Hur72]. In an FM material, there is an additional contribution to the Hall voltage known as the anomalous or extraordinary Hall effect [59, 25, 60, 61, 62, 63]. This voltage difference contribution depends on the value of material magnetisation, and in some materials, it can be larger than the ordinary Hall voltage [59, 60]. Since the strength of the anomalous Hall effect is a function of spin-dependent scattering of the charge carriers, a change in the magnetic state and electrical conductivity of the sample results in a change in the total Hall voltage [59, 25, 60, 61, 63].

As already discussed in Section 2.3, these physical processes open up opportunities to link the concept of magneto-electronic hydrogen gas sensors with the existing electronic gas detection technologies that, in particular, enable a simultaneous measurement of two independent parameters affected by $H_2$ – resistivity and magnetisation. Yet, using the anomalous Hall effect one can create a magneto-electronic sensor that can detect two gases – $H_2$ and CO [64].

It is well-established that hydrogen is highly soluble in Pd. This process results in a significant expansion of the lattice of Pd upon absorption of $H_2$ (0.15% in the α-phase and 3.4% in the β-phase) also leading to an increase in the electrical resistivity of Pd due to the conversion into palladium hydride [84]. A similar behaviour has also been observed in Pd-based alloys [85]. Earlier studies of CoPd alloys and multilayers also revealed a strong dependence of the magnitude and polarity of the AHE signal on the volume concentration of Co [86]. Those findings motivated further research on alloyed structures since it has been hypothesised that absorption of hydrogen by Pd would modify the electronic state of the thin-film structure, thereby affecting the AHE signal and opening up new opportunities for $H_2$ gas sensing [15, 87, 88].

To verify these assumptions, in Ref. [15] polycrystalline $Co_xPd_{1-x}$ films with the atomic concentration of Co in the $0 < x < 0.4$ range were deposited using an e-beam co-evaporation technique. (Note that Co and Pd are completely soluble and form an equilibrium fcc solid-solution phase at all compositions [89, 90].) Several samples with the film thicknesses from 5 to 20 nm were fabricated and characterised without using any post-deposition technique.

For example, Fig. 7 shows the experimental field-dependent hysteresis loops obtained using a 5-nanometre-thick $Co_{0.17}Pd_{0.83}$ film in the $H_2/N_2$ atmosphere at different hydrogen concentrations between 0% and 4%. Significantly, measurements of the films of different thicknesses revealed that thinner films were more suitable for applications in magneto-electronic sensing due to their higher surface-to-volume ratio and since the absolute value of the measured signal is higher for thinner films. The feasibility of the AHE-based magneto-electronic sensors was further confirmed by a demonstration of the detection of low concentration of $H_2$, where it was shown that the sensitivity of the sensor could exceed 240% per 104 ppm at $H_2$ concentrations below 0.5% in the $N_2/H_2$ atmosphere [15]. Overall, those encouraging results indicate that a more than two orders of magnitude higher sensitivity compared with the competing resistivity-based sensor architectures could be achieved using the AHE-based sensor state reading mechanism.

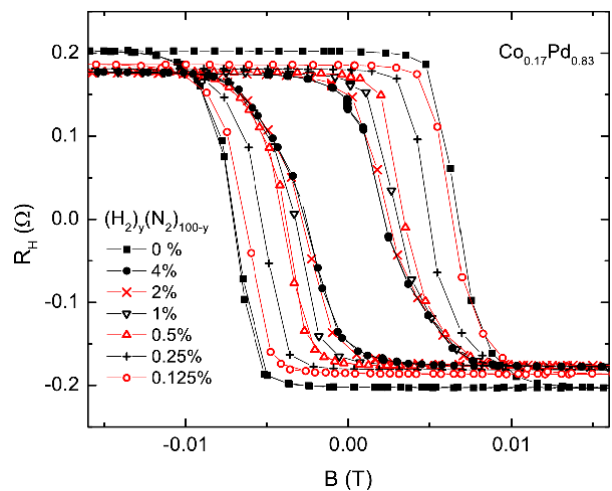

**FIG. 7**: Anomalous Hall effect (AHE) resistance hysteresis loops for a 5-nanometre-thick $Co_{0.17}Pd_{0.83}$ sample in the $H_2/N_2$ atmosphere at different hydrogen concentrations: $y = 0\%$, 0.125%,

0.25%, 0.5%, 1%, 2% and 4%. Reprinted from [15], with the permission of AIP Publishing.

Alloying has also been known as an alternative method for improving catalytic activity and increasing the capability of hydrogen storage systems. In fact, it has been shown that in these contexts Pd-based bimetallic alloys can outperform pure Pd [91, 92]. Therefore, a CoPd alloy was also used to enable the detection of a second gas – CO – alongside $H_2$ using AHE as the principle mechanism of the registration of the state of the sensor [64]. In that work, $Co_{20}Pd_{80}$ alloy films with a fixed thickness of 40 nm were fabricated using magnetron sputtering and their selectivity to $H_2$ and CO was established using a combination of MOKE and AHE. In particular, in the AHE measurements, data were collected with the rate of one data point per second using an Arduino electronic prototyping platform including a WiFi module (see Section 8 for this discussion). A reference multilayer Co/Pd sample was also investigated using the same experimental protocol to demonstrate the advantages of the alloy-based structure.

The hydrogenation of both the alloy film and the reference multilayered sample resulted in a reversible change in the saturation magnetisation. However, whereas the multilayered sample was completely insensitive to CO, the alloy sample exhibited a significant and reversible change in the saturation magnetisation when both $H_2$ and CO were present in the test gas cell. The hydrogen-induced changes in the saturation magnetisation are due to the insertion of hydrogen into the CoPd lattice, which is a well-established fact for both multilayers [44] and alloy structures [17]. This effect results in a local distortion of the lattice and modification of the magnetic order. However, CO with a molecular size greater than that of hydrogen is unlikely to cause a similar effect, and thus it tends to produce magnetic changes only through surface modifications that are possible mostly in the alloy sample.

Figure 8(a) shows a sketch of the experimental setup employed to measure the Hall resistivity in the vertical ($R_{xy}$) direction of the device plane. Figures 8(b.i-b.ii) show the behaviour of $R_{xy}$ of the multilayer sample exposed to $H_2$ and CO, respectively, where one can clearly observe an AHE response due to the exposure to $H_2$ but no changes due to exposure to CO. On the contrary, the behaviour of $R_{xy}$ of the alloy sample exposed both to $H_2$ and CO [Figs. 8(b.iii–b.iv)] clearly shows a shrinkage of the AHE-registered magnetic hysteresis loops.

Furthermore, Figs. 8(c.i-c.ii) show the absolute value of $R_{xy}$ obtained for various $H_2$/CO concentrations as a function of time for the multilayer and alloy sample, respectively, where one can see that both samples exhibit a reproducible $H_2$-induced response in time but only the alloy sample is sensitive to CO. Finally, Figs. 8(c.iii-c.iv) show the concentration dependence of the absolute value of $R_{xy}$ for the multilayer and alloy samples, where one can see that the multilayer structure performs better in a scenario of high $H_2$ concentrations but the alloy structure appears to be more suitable for the detection of low concentrations of $H_2$/CO.

Thus, those encouraging results demonstrate that alloy-based magneto-electronic hydrogen sensors using the AHE detection mechanism hold the potential to find a niche in real-life applications. However, additional optimisation is still required to develop a feasible sensor capable of operating in a wide range of $H_2$ gas concentrations.

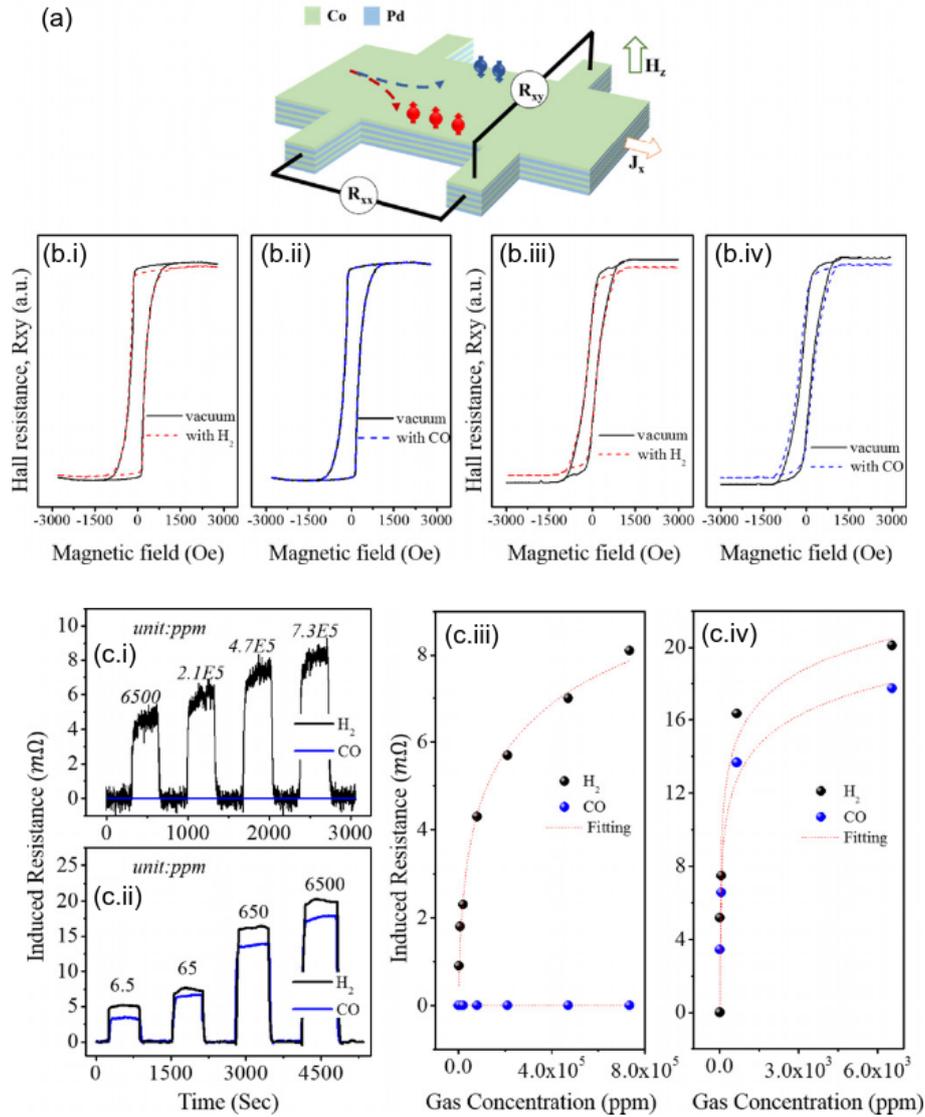

**FIG. 8**: **(a)** Sketch of the AHE-based experimental setup. **(b)** Hall resistance $R_{xy}$ measured in an $H_2$ and CO atmospheres using (b.i-b.ii) the multilayer sample and (b.iii-b.iv) the alloy sample. **(c)** Time-dependent values of $|R_{xy}|$ at various $H_2$/CO concentrations for (c.i) the multilayer sample and (c.ii) the alloy sample. The respective concentration dependencies are shown in the panels (c.iii) and (c.iv). Reprinted with permission from [64].

## 5. Hydrogen sensitivity of the FMR response of PdCo alloyed films

Although the results presented in the seminal work on FMR-based magneto-electronic sensors [13] already demonstrated some significant advantages of such sensors over their competitors (Section 2.1), it was established that further improvements would be needed to fulfil the technical requirements of real-life applications. Indeed, a feasible magneto-electronic sensor should detect the presence of trace (ppm) amounts of $H_2$, and a concentration meter based on the magneto-electronic hydrogen detection mechanism should resolve difference in $H_2$ concentrations from 10% to 100%.

To check whether alloy thin films can help to meet those strict specifications, the FMR response of $Co_xPd_{1-x}$ alloy-based sensors with a varying material composition parameter $x$ ($x$ = 0.65, 0.39, 0.24 and 0.14) was investigated in the work [22]. Figures 9(a, b) show a series of raw FMR traces for the alloy samples taken for a number of different $H_2$ gas concentrations. An analysis of these data revealed that the hydrogen-induced peak shift (in the following we denote it $\delta H$) is much bigger for the sample with $x$ = 0.24 than for the $x$ = 0.39 sample for the same $H_2$ gas concentrations. These data also demonstrate that $\delta H$ exceeds 10 times the linewidth for the highest $H_2$ concentration and it remains larger than the linewidth $\Delta H$ at the lowest concentration of 0.1%. However, for the $x = 0.39$ sample $\delta H < \Delta H$.

Furthermore, one can observe that the FMR response amplitude for the sample with $x = 0.24$ can exceed 30 mV, which is six times larger than 5 mV observed in the bilayer Pd/Co film with the Co layer thickness of 5 nm [13] that we use in this discussion as a reference. However, the concentration of Co in this particular alloy is just 24% of that in the reference bilayer film. Hence, accounting for the three times larger thickness (15 nm) of the alloyed film, the FMR response per Co atom was estimated to be approximately 8.3 times larger for the alloy sample than for the reference bilayer film (potential differences in crystal lattices for the two materials were neglected in this analysis).

The absorption amplitude for the sample with $x = 0.39$ is about 6 mV and thus the Co content in it is approximately 1.5 times larger than for the sample with $x = 0.24$. Therefore, relative to the reference bilayer film, the FMR absorption amplitude per one Co atom for the alloyed film is approximately 1.1, i.e. eight times smaller than for the $x = 0.24$ sample. However, the $x = 0.39$ sample is characterised by an approximately 2.75 times larger linewidth $\Delta H$. Thus, since for the same amplitude of the microwave driving field the FMR amplitude scales as $1/\Delta H$, if the value of $\Delta H$ for this sample were the same as for one with $x = 0.24$ then the absorption amplitude for it would be three times larger than for the reference bilayer film. These calculations indicate that reducing $\Delta H$ for the $x = 0.39$ alloy to the level of $x = 0.24$ could be beneficial for magneto-electronic hydrogen gas sensing.

To complement the current discussion, we discuss the response time of the sample with $x = 0.24$ to exposure to $H_2$ gas as well as the time required by the respective FMR signal to return to its original state after evacuation of $H_2$ from the test gas cell [Fig. 9(c)]. That measurement was conducted using a time-resolved FMR method [39], where the frequency was set to be 10 GHz and the bias magnetic field was 4160 Oe. The dashed line in Fig. 9(c) shows the concentration of $H_2$ as a function of time. Initially, the sample environment was pure $N_2$. At 480 s into the experiment, the atmosphere was replaced with one containing 0.1% $H_2$. At 1800 s, the test gas cell was flushed with pure $N_2$ gas that was then kept flowing until the end of the measurement. Consistently with the FMR results in Figs. 9(a, b), the change in the FMR amplitude caused by $H_2$ gas entering the test cell is very large (30 mV). Although there is a sharp response of the film to the presence of $H_2$ gas evidenced by a very steep slope of the ascending section of the curve, the recovery of the system is slow since the descending section of the curve is much flatter.

Continuing the discussion of the FMR traces, we note that in practical magneto-electronic hydrogen gas sensors, a material with a large value of $\Delta H$ and a relatively strong FMR response, such as 6 mV observed for $x = 0.39$, could also be useful. As can be seen from Fig. 3(b), the FMR peak shift $\delta H$ for this sample in the presence of just 0.1% $H_2$ is large and easily detectable. Accordingly, the change in the amplitude of the FMR signal once $H_2$ has been let into the test gas cell is very large as well (30 mV). At the same time, it was also observed from the time-resolved FMR measurements that the response and recovery for the sample with $x = 0.39$ is much faster than for $x = 0.24$.

Significantly, previously in Ref. [39] it has been shown that it is possible to cover a wide range of $H_2$ gas concentrations using a single sensor device by tuning its sensitivity to different concentration sub-ranges. Such a tuning is possible by adjusting a static magnetic field applied to the film and it is needed because $\delta H > \Delta H$. However, no field adjustment would be needed if the constituent material of the sensor can produce the same FMR amplitude and the same value of $\delta H$, also having $\Delta H$ that is equal or larger than the maximum value of $\delta H$. The sample with $x = 0.39$ can be such a material.

Thus, to summarise the discussion above, it has been demonstrated that the alloy-based magneto-electronic sensors employing FMR can measure $H_2$ concentration in the range from 0.05% to 100%. Furthermore, as shown in Fig. 9(d), the concentration resolution of the alloy-based sensors is non-vanishing up to 100% concentration and it scales approximately as $1/C$, where $C$ is the concentration of $H_2$. Significantly, it was also demonstrated that the alloy-based sensors exhibit no saturation near the 100% mark. Such a behaviour of the FMR-based sensor is in stark contrast with that of the entire group of Pd-based solid-state sensors that saturate at relatively low $H_2$ concentrations [21]. As a result, alloy-based magneto-electronic sensors are projected to be suitable for applications as $H_2$ concentration meters operating inside a fuel cell [5].

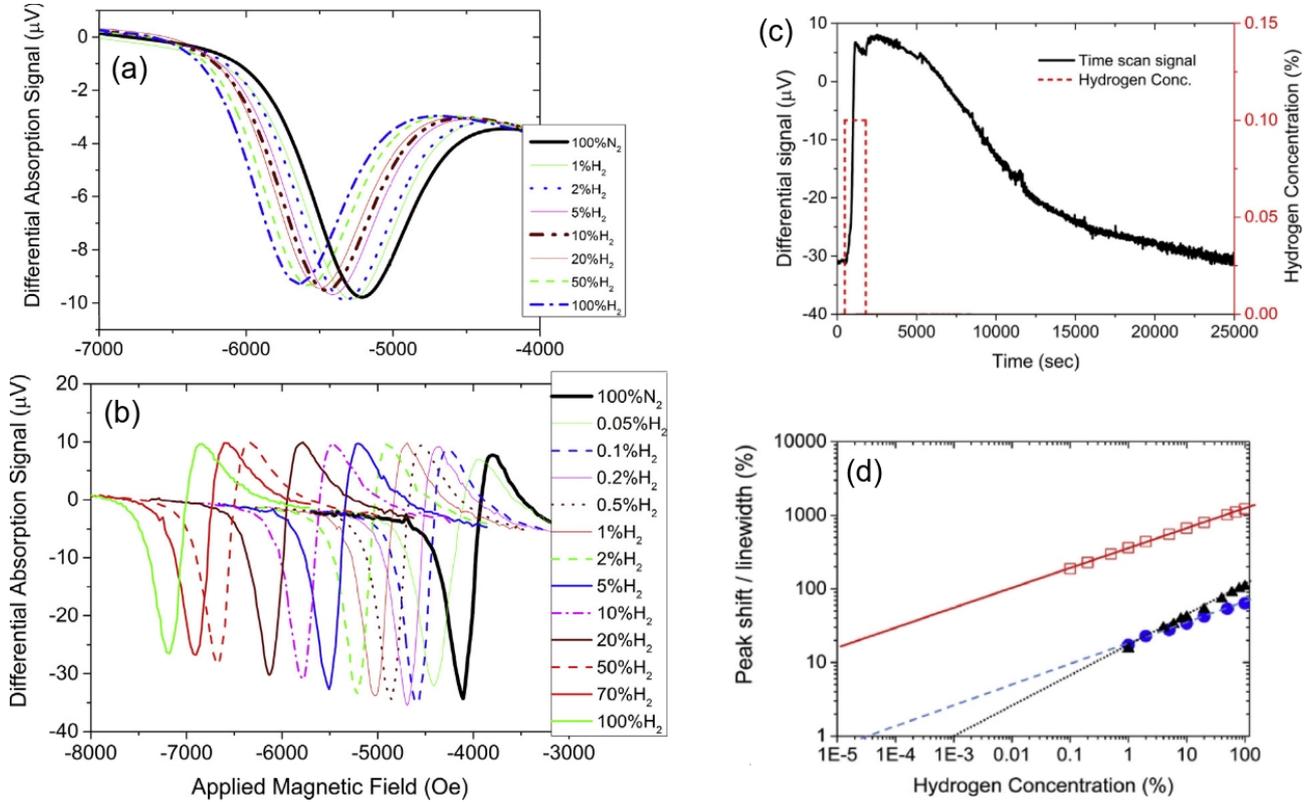

**FIG. 9**: Representative FMR traces produced by $Co_xPd_{1-x}$ alloy films for different concentrations of $H_2$ gas: (a) $x = 0.39$ and (b) $x = 0.24$. Measurements are taken at the frequency of 10 GHz. (c) Response and recovery times for $Co_xPd_{1-x}$ alloyed films with $x = 0.24$. Hydrogen concentration in the gas mix is 0.1%. (d) Response of $Co_xPd_{1-x}$ alloy films with $x = 0.24$ (squares) and $x = 0.39$ (circles) compared with the response to $H_2$ gas of a high-performing reference Co/Pd bi-layer film-based sensor (triangles). Data extracted from FMR measurements are plotted on a logarithmic scale and the straight lines are the linear fits to the experimental points. Reproduced with permission from [22].

Another important outcome of the measurements conducted in Ref. [22] is the demonstration of a significant variation of the FMR peak linewidth $\Delta H$. In general, one may expect that the resonance line for Co would be increased when one dilutes Co with non-magnetic Pd. Indeed, the alloyed films fabricated in Ref. [22] exhibited a broad range of values of $\Delta H$. As we already mentioned, in an FMR experiment, there is an important relation between $\Delta H$ and the height of the FMR peak: the wider the peak, the smaller its height. On the other hand, technically, the larger the peak height, the simpler electronics needed to detect the peak, and hence, the smaller the fabrication costs of a future FMR-based hydrogen sensor. Thus, there is a strong incentive to maximise the strength of FMR response.

The height of the FMR peak also scales as the volume of resonating material. Above we mentioned that the effect of the interface PMA on the FMR peak position is significant only for thin magnetic layers. For that reason, the Pd/Co bilayer based hydrogen-gas sensor prototypes had a Co layer with the thickness of 5 nm [13]. In that case, the FMR response peak was small but still detectable with available equipment, but the observed $H_2$-induced peak shift $\delta H$ was sufficient to reliably detect $H_2$ concentrations as small as 0.06%.

The Pd alloy samples possess bulk PMA and, therefore, they do not need to be as thin as bilayer films to produce a significant value of $\delta H$. Hence, there is a possibility to compensate the decrease in the height of the FMR peak due to an increased $\Delta H$ by using a thicker alloy film. The experiment from Ref. [22] also confirms this idea: a 15 nm thick film with a broad linewidth $\Delta H$ = 700 Oe was used there and the FMR peak height was found to be similar to the one for a reference Pd(10 nm)/Co(5 nm) bilayer film.

As we already discussed, alloy samples that exhibited a large $\Delta H$ would be particularly suitable for FMR-amplitude-based $H_2$ sensing, where, in contrast to Co/Pd bilayer films [24, 39], there would be no need to re-adjust the sensor several times to conduct measurements in the same broad $H_2$ concentration range. In this context, we note that it would be challenging to artificially increase $\Delta H$ of a layered sample with the "standard" thickness of Co layer of 5 nm to enable the same adjustment-free operation since in that case the FMR peak would become too weak to reliably detect it on top of the noise background. In other words, using thicker alloyed samples one can increase $\Delta H$ without compromising the ability to detect the FMR response itself, and thus enabling a re-adjustment-free measurement of a broad range of $H_2$ concentrations. However, the CoPd alloy films with a significantly smaller value of $\Delta H$ could also find applications since their FMR properties would be ideally suitable for frequency-based hydrogen sensing, where the $H_2$ concentration would be effectively encoded in the FMR frequency shift caused by the hydrogenation.

Finally, recall that FMR does not distinguish between a change in the saturation magnetisation and a change in PMA [see the discussion around Eqs. (1-3)]. As follows from the raw FMR traces [Figs. 9(a,b)], the magnitude of FMR field for the employed OOP FMR configuration decreases with an increase the $H_2$ concentration. This may be due to either a $H_2$-induced decrease in PMA or an increase in $4\pi M_s$. Which exactly of these effects takes place could not be determined in Ref. [22] since doing so would require measurements with an alternative method that can separate these parameters (e.g. Vibrating Sample Magnetometry). Moreover, those measurements have to be conducted in the presence of hydrogen gas, which is technically challenging and potentially unsafe.

## 6. Effect of hydrogen gas on FMR properties of NiCoPd ternary alloy films.

Nickel (Ni) is often employed in Pd-based hydrogen gas sensors that operate using the physical processes resulting in variations of the electric resistivity and optical response of the active material. Apart from playing a role in corrosion protection and increased mechanical strength of active layers of gas sensors [74, 20, 93], the addition of Ni accelerates the processes of absorption and desorption of $H_2$. Yet, the addition of Ni helps prevent a failure of a sensor exposed to more than 20% of $H_2$ in the environment, when an irreversible saturation of the electric-resistivity response can occur due to a transition from the α-phase of Pd to the β-phase [19, 20, 74, 93]. Furthermore, the presence of Ni yields a contraction of the Pd lattice [94], which in bulk samples of PdNi alloys results in a decrease in the solubility of hydrogen and a suppression of the phase transition to the β-phase. As a result, in PdNi alloys with 8-15% of Ni, significantly higher $H_2$ gas concentrations are required to trigger a saturation [19].

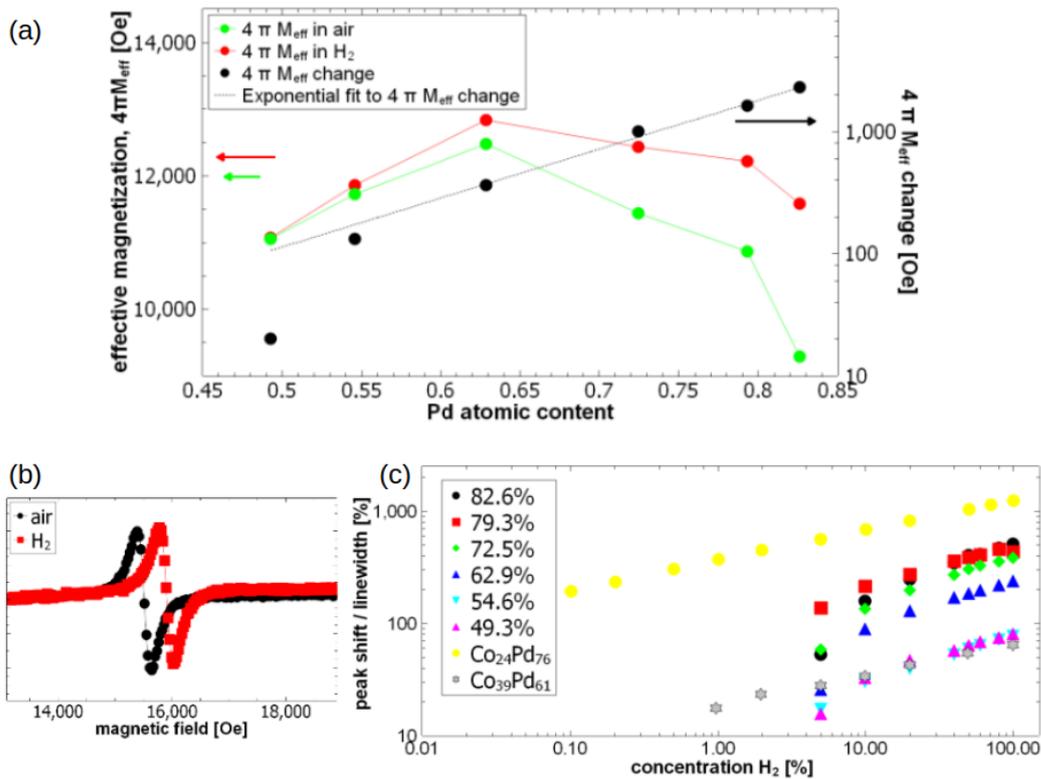

**FIG. 10**: **(a)** Left axis: effective magnetisation $4\pi M_{eff}$ in air (the green dots) and in $H_2$ atmosphere (the red dots) as a function of the Pd content in the material. Right axis: difference in $4\pi M_{eff}$ plotted on a logarithmic scale (the black dots) and an exponential fit to it (the black dashed line). **(b)** FMR response at 9 GHz for the $Ni_{26.1}Co_{11.0}Pd_{62.9}$ sample in $H_2$ environment (the red squares) and in air (the black dots). **(c)** Normalised FMR peak shift for the NiCoPd alloy films extracted from the FMR data as a function of the $H_2$ concentration. The numbers from 49.3% to 82.6% denote the Pd content in the NiCoPd alloy films. The data '$Co_{24}Pd_{76}$' and '$Co_{39}Pd_{61}$' are for the respective reference CoPd alloy films. Reproduced with permission from [95].

Since the FMR spectroscopy is used to register the state of the magneto-electronic sensors discussed in this section, it is worth noting that Ni is an FM transition metal that is characterised by the smallest saturation magnetisation and the largest FMR linewidth $\Delta H$ compared with Co and Fe. Therefore, apart from playing several roles in the detection of hydrogen as discussed above, the addition of Ni to the sensitive material of a magneto-electronic sensor will always additionally affect the fundamental physics of the FMR response of the sensor.

Given the aforementioned general advantages of use of Ni in various gas sensors, it is plausible to assume that the addition of Ni to Co- and Pd-based structures of magneto-electronic sensors would help improve the ability of such sensors to detect $H_2$. This hypothesis was verified in the work [95], where a number of NiCoPd alloy films were fabricated and investigated using the FMR spectroscopy in air and in $H_2$ environments. Overall, a behaviour similar to that of CoPd alloy thin films was observed at high concentrations of $H_2$. However, while the response of CoPd thin films steadily increased as the $H_2$ concentration was reduced, for lower $H_2$ concentrations from 0 to 5% the investigated NiCoPd samples demonstrated mostly a sudden decrease in the sensitivity. Moreover, when compared with the response of CoPd films, the NiCoPd films produced a lower normalised FMR peak shift at a higher Pd content and a larger peak shift at a lower Pd content.

Such a behaviour was attributed to the certain differences in the fundamental physical processes occurring in the constituent materials of the alloy films. To demonstrate this, the effective magnetisation $4\pi M_{eff}$ and the gyromagnetic ratio $\gamma$ for the samples were extracted by fitting the FMR traces with the OOP Kittel formula Eq. (3).

It was established that the dependence of $4\pi M_{eff}$ on the Pd content is not linear but first it exhibits an increase and then a decrease [Fig. 10(a)]. An increase in $4\pi M_{eff}$ was previously observed in similar NiFePd alloys [96], where it had been attributed to an increase in the concentration of FM species (Ni and Co) that effectively increases the saturation magnetisation of the material. However, the decrease in $4\pi M_{eff}$ at around 63% of Pd content in Fig. 10(a) was attributed to an effective decrease in the magnetic moment of Co caused by a lower magnetic moment of Ni. Yet, in close proximity to FM atoms the atoms of Pd become magnetically polarised thereby contributing to the effective magnetisation $4\pi M_{eff}$. This effect is weaker in a NiPd alloy than in a CoPd one [97].

As seen from Fig.10(b), the strength of alloy response to hydrogen loading while expressed in terms of a change in $4\pi M_{eff}$ grows exponentially with Pd content in the alloy. Several physical factors were identified to be responsible for this behaviour:

(1) the absorption of $H_2$ modifies the electronic structure of a Co/Pd layer, thereby changing the magnetic environment of the atoms and reducing the strength of PMA [22, 98]. A plausible explanation of this effect is a change in the Co-Pd orbital hybridisation [99] caused by hydrogen ions [24]. Since the addition of a small amount of Ni should not impact this physical mechanism noticeably, it was concluded that in the NiCoPd alloy structures the change in $4\pi M_{eff}$ should also be due to a decrease in the strength of PMA.

(2) the dilution of Pd with a transition metal reduces the ability of Pd to incorporate hydrogen into its crystal lattice [70]. In turn, a lower concentration of hydrogen in the crystal lattice has a smaller effect on the hybridisation of orbitals of Co and Pd atoms. Yet, a reduction of Pd content also implies that there would be fewer Co-Pd bonds that contribute to PMA [100]. As a result, one can expect a smaller hydrogen-induced reduction of PMA at a smaller Pd content.

In Fig. 10(c) the hydrogen-induced FMR peak shift is normalised to the respective value of $\Delta H$ and is plotted as a function of the concentration of $H_2$ to demonstrate that a higher Pd content in the alloy results in a more pronounced shift in the resonance field $\delta H$. Significantly, this result holds even when the ratio $\delta H/\Delta H$ is analysed, and only at the highest and lowest concentrations of $H_2$ considered in the experiment the effect of a large $\Delta H$ compensates for a larger value of $\delta H$ for the films with a high content of Pd.

Interestingly, in Fig. 10(c) the data points for the NiCoPd films deviate downward with respect to a straight line but the dependence of the normalised peak shift $\delta H/\Delta H$ on the concentration of $H_2$ in CoPd is linear. Thus, at present there is no clear and unambiguous picture of the impact of Ni doping on the hydrogen sensing properties of PdCo alloys. Therefore, additional investigations are needed to gain complete understanding of advantages and disadvantages of more complex ternary-alloy systems.

**7. FMR response of hydrogenated Co/Pd multilayers consisting of ultra-thin films: Relevance to alloyed magneto-electronic hydrogen gas sensors**

In this section, we discuss the results of a work, where the property of an interface between Co and Pd layers was employed to fabricate a "pseudo-alloy" – a material with signatures of both being an alloy of the two metals and being a sequence of layers with magnetic interfaces between them at the same time.

Let us recall that in multilayered FMR-based magneto-electronic sensors the effect of PMA is usually observed at an interface between a Co layer and a Pd layer, and that a change in the strength of PMA caused by exposure of the multilayered structure to $H_2$ gas is often registered as a shift $\delta H$ in the position of a resonance peak. Whereas the absolute value of this shift is inversely proportional to the thickness of the FM metal layer, the amplitude of the FMR signal scales with the thickness of the FM layer. Subsequently, there is a trade-off between the amplitude and the PMA-mediated resonance shift. On the other hand, using CoPd alloy films instead of Co/Pd multilayers allows observing an FMR response that is free of such a trade-off since the behaviour of alloyed films is governed by bulk PMA. However, while this property of alloys is advantageous for practical applications, using it comes at a cost: fabrication of thin films made of alloys of arbitrary composition may pose several technological challenges compared with well-established techniques used to produce thin-film multilayers.

This situation has motivated research on layered structures containing a region of alternating ultra-thin Co/Pd films that combine certain characteristics of both conventional multilayers and alloyed films, thereby allowing one to circumvent the trade-off between the magnetic layer thickness and the effect of hydrogenation on PMA while keeping the fabrication process simple. This section summarises the main experimental results obtained in a relevant recent work Ref. [101], where polycrystalline multilayer thin films with the nominal layer structure Pd(6 nm)/[Co(0.19 nm)/Pd($x$ nm)]$_{30}$/Pd(6 nm) were fabricated using a magnetron sputtering system and characterised in air and in $H_2$ gas atmospheres using FMR spectroscopy [Fig. 11(a)].

It has been demonstrated that exposure of such samples to $H_2$ leads to an upward shift $\delta H$ in the FMR field, which is an expected results, and that the origin of this shift is due to a change in the effective PMA field that enters the expression for the effective magnetisation: $4\pi M_{eff} = M_S - H_{PMA}$. In turn, the change in PMA can be due to both hydrogen-induced mechanical strain in Co resulting in a magnetostriction contribution to the bulk anisotropy of Co layers and hydrogen ion induced electronic effect on the strength of the Co-Pd bond [24].

To reveal the actual composition of the fabricated samples compared with the nominal one, the cross-sections of the films were comprehensively characterised using a high-resolution transmission electron microscopy (TEM) and a scanning transmission electron microscopy with energy dispersive X-ray spectroscopy (STEM/EDX). A theoretical Kittel equation-based model [Eqs. (1-3)] was employed to analyse raw FMR spectroscopy data and also to reveal a dependence of the effective magnetisation $4\pi M_{eff} = 4\pi M_s - H_{PMA}$ on the ratio of Co/Pd layer thicknesses $t_{Co}/t_{Pd}$ [Fig. 11(b)].

In each investigated sample, TEM and STEM/EDX measurements demonstrated a homogeneous mixture of Co and Pd with an averaged composition varying from $Co_{0.33}Pd_{0.67}$ to $Co_{0.5}Pd_{0.5}$. No indication of layering was observed. Since the spatial resolution of STEM/EDS is, in general, insufficient to fully resolve the layer thicknesses of the investigated structures, X-ray reflectivity (XRR) measurements were also conducted since XRR can resolve subnanometre-thin layers provided the interface between them is abrupt. However, XRR measurements did not reveal the presence of any well-defined periodic multilayer structure either. All those observation are consistent with the fact that fabrication of well-defined ultra-thin layers of Co/Pd is technically challenging and that a particular magnetron sputtering regime used to fabricate the investigated multilayers is known to promote island-like polycrystalline film growth. Moreover, the growth of the films could be affected by intermixing of Co with Pd [99]. On the other hand, the possibility of alloying due to atom migration was ruled out since the sputtering temperature (180 ºC) was not high enough to trigger this physical process. Therefore, it was concluded that the island-like growth of the discontinuous individual layers promoted the creation of an alloy-like structure with no clear layering, but nevertheless the samples retained certain properties of thicker well-defined multilayers.

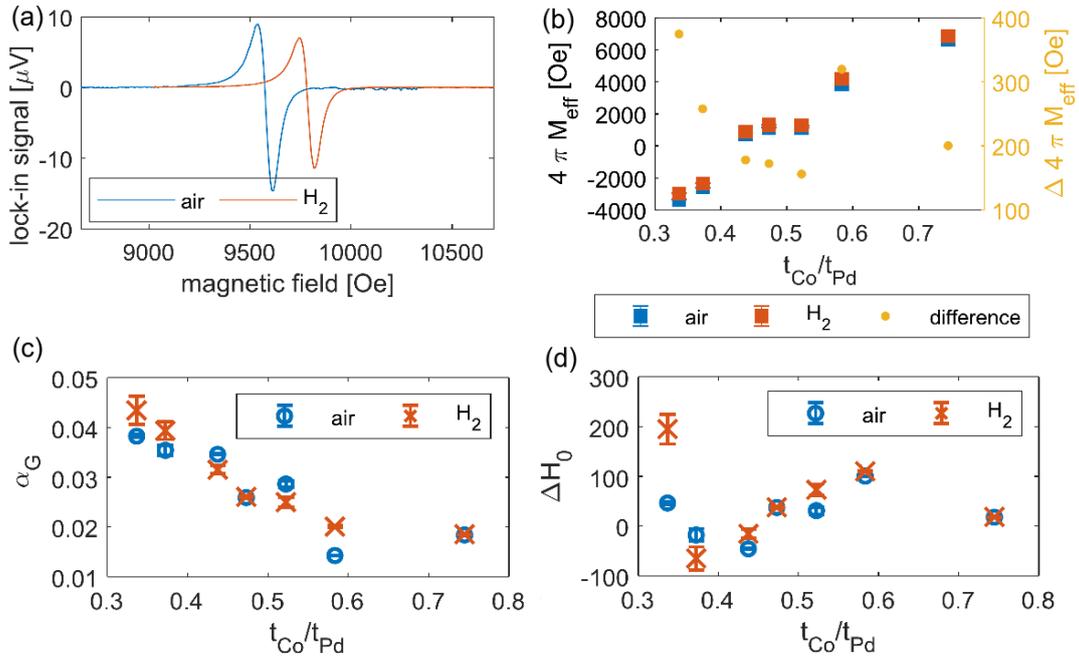

**FIG. 11**: **(a)** Representative FMR signals produced by an ultra-thin-film Co/Pd sample in air and $H_2$ atmospheres. **(b)** Values of the effective magnetisation $4\pi M_{eff}$ extracted from raw FMR data using a Kittel equation-based model in air and $H_2$ atmospheres (left axis). The difference in the value of $4\pi M_{eff}$ extracted from the measurements conducted in air and $H_2$ are plotted in the right axis. **(c)** Gilbert damping constant $\alpha_G$ and **(d)** inhomogeneous linewidth broadening $\Delta H_0$ extracted from the experimental values of FMR linewidth. Reproduced with permission from [101].

The effective magnetisation $4\pi M_{eff}$ extracted from the the Kittel equation-based model increases with an increase of the $t_{Co}/t_{Pd}$ ratio [Fig. 11(b)]. This result agrees with a previous observation made in similar structures [102], where it has been demonstrated that a change from an easy perpendicular magnetic axis to a hard one takes place when the $t_{Co}/t_{Pd}$ ratio is increased. Furthermore, compared with CoPd alloy thin films of a similar composition [22], in the current samples the value of $4\pi M_{eff}$ depends much stronger on the Co content. For instance, in Ref. [22] a $Co_{0.14}Pd_{0.84}$ sample exhibited a positive value of $4\pi M_{eff}$ compared with a negative value of the effective magnetisation obtained for the ultra-thin-film sample with the lowest Co content. Such a difference has been attributed to the existence of some residual interfacial processes that may be undetectable using XRR or electron microscopy but that still can induce an interfacial PMA [103].

Thus, since the currently discussed samples contain alloy layers but PMA is induced by the residual interfaces due to the growth process, one may expect that several contributions of PMA to the response of the sample would be observed. In fact, as can be seen from Fig. 11(b), the behaviour of $4\pi M_{eff}$ scales linearly with the $t_{Co}/t_{Pd}$ ratio. However, the respective FMR peak shift caused by a change in PMA due to hydrogenation increases sharply for a low $t_{Co}/t_{Pd}$ ratio. A possible explanation of this result is that there are at least two sources of PMA in the Co/Pd ultra-thin-film multilayers: a bulk PMA and an interface PMA. The contributions of both kinds of PMA increase with an increase in Pd layer thickness as evidenced by a reduction in the $t_{Co}/t_{Pd}$ ratio. The bulk PMA may originate from the fact that, similarly to CoPd alloys, the Co layer was growing in a Pd-like crystal lattice. Yet, as established in Ref. [42], a Co/Pd multilayer does not expand upon hydrogen absorption, thereby indicating that the bulk PMA would be unaffected by the absorption while the interface PMA would be changed.

As a next step, the linewidth $\Delta H$ of the FMR resonance line was extracted by fitting the field-resolved FMR traces with a standard model equation. Then the $\Delta H$ values were converted into values of the Gilbert damping constant $\alpha_G$. While the values of $\Delta H$ typically showed a higher measurement uncertainty compared with the extracted values of the resonance field, $\alpha_G$ demonstrated a clear trend with the increase in the $t_{Co}/t_{Pd}$ ratio. This result is in good agreement with the previous relevant observations of enhanced damping of the magnetisation precession due to spin pumping [104, 105].

The inhomogeneous linewidth broadening $\Delta H_0$ is related to the linewidth $\Delta H$ and it describes the impact of extrinsic damping effects that include inputs from local variations of the magnetisation and anisotropy constants. In general, the observed behaviour of $\Delta H_0$ was mostly random but for several $t_{Co}/t_{Pd}$ ratios even negative values of $\Delta H_0$ were obtained. While a negative value of $\Delta H_0$ is unphysical, an observation of it may indicate that (i) the relationship between the frequency $f$ and $\Delta H$ is not linear, which might be the case of a film that is not fully saturated, or (ii) the values of $\Delta H$ are significantly scattered. Nevertheless, for $t_{Co}/t_{Pd} = 0.34$, a significant increase in $\Delta H_0$ was observed when the sample was hydrogenated and that increase was well-above any error that could arise during the fitting of data. That observation was explained by the fact that the presence of Pd could result in exchange-coupling between different regions of the Co islands that are characterised by different saturation magnetisation and magnetic anisotropy. Hence, it was plausible to assume that hydrogenation might weaken this exchange interaction if not completely destroy it, thereby resulting in a frequency-independent increase in $\Delta H$. A number of previous observations speak in favour of this assumption, most notably the facts that Co/Pd multilayers exhibit a small change in the coercive field upon hydrogenation [42, 43] but a strong increase in the coercive fields was observed in $Co_{14}Pd_{86}$ alloy thin films at any partial $H_2$ pressure [72].

Finally, in general, $\Delta H$ did not show any significant hydrogen-induced variation (apart from the particular sample discussed just above), which is in contrast to the behaviour of Co/Pd bilayers, where $\Delta H$ consistently reduced as a result of absorption of $H_2$. Such a reduction was explained by a decrease in the strength of spin-pumping [95] and it was assumed that due to the thicker Pd layers this process might be more pronounced for bilayers than for multilayers. However, in multilayers made of FM and rare-earth metals a reduction in $\Delta H$ due to hydrogen exposure has also been attributed to the interface clearing effect. Interpreting those results in the context of the structures discussed in this section, the interface clearing effect may result in a reversible improvement of structural quality of the interface between Pd and an FM metal. It has been suggested in [47] that in the presence of hydrogen the atoms of the interface can become more mobile and hence they can rearrange themselves into a more regular interface. Therefore, scattering of the FMR precession mode from interface inhomogeneities is reduced and this results in a smaller value of $\Delta H$ and a large amplitude of the FMR peak. This effect is reversible, which means that the interface returns to its original state when hydrogen is removed from the material. While there is no any evidence in support of the presence of this process in the Co/Pd multilayers discussed in this section, one cannot exclude its contribution to $H_2$-induced changes in $\Delta H$ for the thicker Co/Pd bilayers. Interestingly, no significant change in $\Delta H$ was observed in the CoPd and NiCoPd alloy thin films discussed above. Thus, it is plausible to assume that ultra-thin-film multilayer samples might actually be more alloy-like than multilayer-like.

## 8. Prototypes of alloy-based magneto-electronic gas sensors

Despite a number of technical and fundamental challenges that the developers of magneto-electronic hydrogen sensors will need to resolve in the future, the magneto-electronic sensors have already reached the stage of device prototyping and intellectual property protection (see, e.g., [106]). Here, we overview some of the results that approach us to an era of commercial magneto-electronic gas sensors and outline potential challenges that we may encounter on this journey.

For example, in Ref. [64] an Arduino electronic prototyping platform was used to enable the interaction of electronic objects of the proposed sensors, such as those controlling the concentration of $H_2$ in experimental lab settings. The entire sensing device consisted of an Arduino platform, a sensitive alloy film-based layer, an amplifier module, an ACD module and a WiFi module (see Fig. 12, where a screenshot of a supplementary video file in Ref. [64] is presented ). In particular, the values of $|R_{xy}|$ (see Fig. 8 and the discussion around it) were digitally converted using one of the modules so that they could be monitored on a graphic display similarly to a commercial gas sensing system.

Once one has a full schematic circuit designed using an Arduino platform, the next step is to turn it into a real-world printed circuit board (PCB). Given the wireless functionality of the sensor prototype, one has to pay special attention to the PCB layout for the antenna to maximise the power transmission and minimise interference with the internal microwave electron devices and systems. In addition, the Arduino code should be ported to a native firmware code for the particular microcontroller models used in the design. However, these technical aspects mostly belong to the area of electronic device design and testing and they are not directly relevant to the physics of magneto-electronic gas sensors. Nevertheless, the design of some of the sensor's elements might need to be changed to facilitate its integration with the electronic modules and especially antennas that can also operate at GHz-range frequencies.

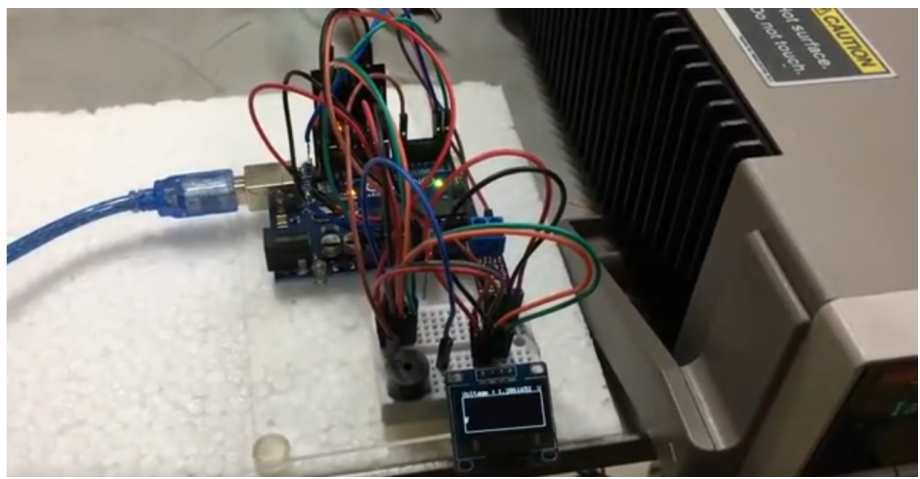

**FIG. 12**: Screenshot taken from a supplementary video from Ref. [64] (https://www.youtube.com/watch?v=7fQ2xwPDQwc&feature=youtu.be) demonstrating an Arduino-platform prototype of a magneto-electronic $H_2$/CO sensor proposed in [64].

Whereas this approach to converting an Arduino prototype into a manufacturable product significantly optimises the process of finding and fixing design flaws, there exist alternative ways of device prototyping that can be more compatible with the equipment available at university-based and industrial R&D centres and that can lead to a faster transition from the prototyping stage to a real-life device. For example, following this approach, an on-chip integrated magneto-electronic hydrogen sensor using the FMR as the detection mechanism (Fig. 13) was proposed in Ref. [107]. Note that that work employed Fe instead Co in the active alloy layer. An FMR peak upshift of about 80 Oe was observed in that work in the presence of $H_2$ gas. However, an approximately 20 Oe linewidth of the FMR response of the Fe-based sensor is much smaller than that of approximately 150 Oe observed in a typical Co-containing structure, which represents a clear advantage for real-life applications since using it significantly simplifies the registration of the signal caused by the presence of $H_2$ gas.

The design of the sensor in [107] also addresses some challenges faced by the previous generations of magneto-electronic sensors, such as insufficient strength of the FMR signal and the use of relatively expensive laboratory-level electronics to reliably detect the FMR response. In particular, to unleash the commercialisation potential of the sensor, miniaturised magneto-electronic sensors

(microchips) were fabricated using a UV-photolithography technique and an electron-beam assisted evaporative system. The resulting waveguide-like structure [Fig. 13(a)] was characterised in a custom-made pressure-tight gas cell equipped with the microscopic microwave probes "Picoprobes®" (GGB Industries) [26]. The fabricated microchip demonstrated a two-times stronger FMR signal compared with that produced by a reference continuous film. Evidently, this is still work in the progress, but this very first result is already promising.

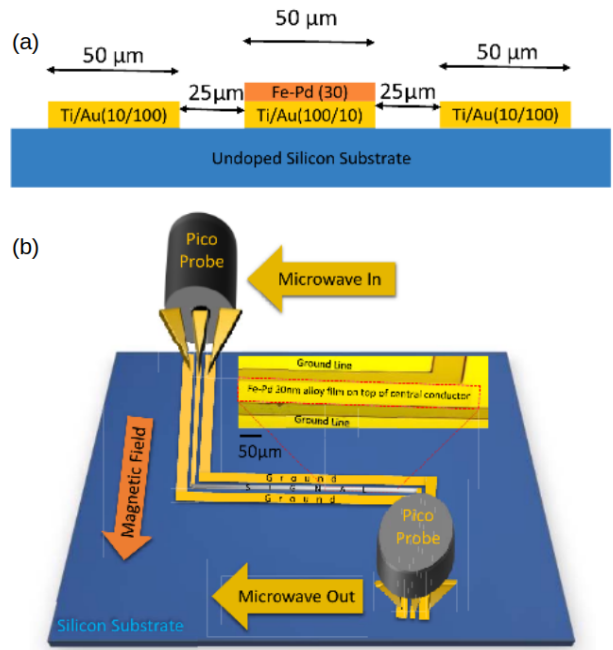

**FIG. 13**: **(a)** Schematic of the cross-section of the fabricated microchip with a coplanar waveguide (CPW)-like configuration. **(b)** Schematic of custom-built FMR characterisation station used to investigate microchips. The inset shows the magnified view of central conductor of the CPW. Reproduced with permission from [107].

## 8. Conclusions and outlook

Thus, in the recent years there have been a significant progress in understanding the fundamental physical processes that underpin the operation of alloy-based magneto-electronic hydrogen gas sensors. It has been demonstrated that, in comparison with their multilayer-film-based counterparts proposed a decade ago ([12, 13], for a comprehensive review see [16]), the alloy-based magneto-electronic sensors exploiting the physical mechanisms of ferromagnetic resonance (FMR), magneto-optical Kerr effect (MOKE) and anomalous Hall effect (AHE) can provide certain advantages for the detection of $H_2$ gas in a wide range of its concentrations. Those fundamental and practical results increase the plausibility of the magneto-electronic hydrogen gas sensing technology as a whole, at the same time clearly indicating a significant potential of the alloy-based magneto-electronic sensors to find a niche on the market of hydrogen gas sensors. Such a positive outlook is supported by the demonstrated ability of several kinds of magneto-electronic sensors to provide a very low fire risk detection of hydrogen gas and therefore to become a technology of choice in mass transport hydrogen-powered vehicles, where the safety is of utmost importance.

However, extra research studies and development work are required for the alloy-based magneto-electronic sensors to reach the industry. Here, we outline some of the important milestones that will need to be accomplished to achieve this goal.

(1) At present, none of the investigated structures demonstrated a combination of all merits of the magneto-electronic hydrogen sensing technology discussed in this review. Firstly, this is because those sensors were fabricated using different sets of constituent materials, and different physical mechanism of the sensor state registration, such as FMR, MOKE, AHE and other effects, have been employed. More specifically, whereas it has been shown that using alloyed active materials may be advantageous in many aspects, for instance, while measuring large $H_2$ concentrations, the actual list of Pd-FM metal alloys that can be used for this application is long but the optimal alloy composition has not been established yet. In addition, in the sub-field of FMR-based magneto-electronic sensors extra work needs to be done to control the FMR peak linewidth in the alloyed films. Similarly, the effect of material nanostructuring on alloy-based structures has not been investigated yet. This situation urges a joint effort of physicists, material scientists, chemists and engineers intended to bridge this knowledge gap.

(2) The research works on magneto-electronic gas sensors have considered mostly thin-film based structures, where the static magnetisation vector naturally lies in the plane of the film. However, structuring thin-film samples by forming, for example, long strips with a square nanoscale cross-section (called nanowires) is known to result in a FMR-based sensor operation at zero applied magnetic field, which is beneficial for real-life applications since it simplifies the design of the sensor [16]. Yet, it was demonstrated that the sensitivity of FMR-based magneto-electronic sensors to $H_2$ can be increased by directing the static magnetic moment of films perpendicular to the film plane. However, in this case, a strong magnetic bias field has to be applied to the material, which currently can be achieved using only large laboratory-level electromagnets that cannot be miniaturised. This problem can be resolved using alloys of Co or Fe with Pd, where the static magnetic moment is reduced so that a low static magnetic field needed to rotate the magnetic moment perpendicular to the plane of the sample can be created using an inexpensive, compact and commercially available permanent magnet. Furthermore, based on purely theoretical considerations, it is plausible to assume that nanostructuring of alloyed materials should decrease the required bias field even further, potentially to zero. It is also clear that in this case the geometry of nanowires will not be suitable since it intrinsically favours an in-plane orientation of the magnetic moment. Thus, other nanopatterning geometries promoting a perpendicular-to-plane orientation of the magnetic moment need to be chosen.

(3) Both MOKE and FMR investigations demonstrated that film nanopatterning helps reduce the time response of multi-layer based sensor towards the industry standard of one second [45, 14]. No similar studies have been conducted for alloy films yet.

(4) It is noteworthy that, since the thickness of the active material films is very small, a relatively high cost of both Co and Pd does not significantly affect the cost of production of a magneto-electronic hydrogen gas sensor. For instance, CoPd alloy layer thicknesses used to generate a measurable FMR, MOKE or AHE response are typically from 5 to 20 nm. It can be estimated that such a magneto-electronic hydrogen gas sensor would typically contain less than 4 μg of Co and less than 2.5 μg of Pd. Given that Pd costs approximately USD 100 per gram and that Pd is much more expensive than Co, the contribution of these quantities of materials to the cost of fabrication of a sensor is rather negligible.

(5) It is known that other gases can contaminate and interfere with Pd-based sensor ability to detect hydrogen. In the magneto-electronic sensors, this effect as well as the effects of humidity, gas flow and a water layer formation on the surface of the sensing element are also likely to be present. However, these effects are essentially the same as for any other pure-Pd and Pd-alloy based sensors, including those based on measuring the conductivity of a Pd-containing layer. Therefore, approaches to the resolution of this and other relevant problems will also be similar. On the other hand, the magneto-electronic sensor concept does not use the effect of hydrogenation on the general physical properties of Pd itself but specifically on the magnetic properties of its alloys with FM metals. Therefore, one may expect that some of the aforementioned adverse effects may affect magneto-electronic sensors in a different way but some of them may even be entirely irrelevant. However, this important aspect of magneto-electronic sensors has not been addressed in the literature yet. Therefore, we believe that the current review would motivate colleagues from academia and industry to bridge this and other existing knowledge gaps.

**Funding**: We would like to thank the support by Australian Nuclear Science and Technology Organisation (ANSTO), grants No. P4123, P4810 and P6126. We also acknowledge using the equipment of and receiving scientific and technical assistance from the Australian National Fabrication Facility and the Centre for Microscopy, Characterisation and Analysis of the University of Western Australia, a facility funded by the University, State and Commonwealth Governments. ISM has been supported by the Australian Research Council through the Future Fellowship (FT180100343) program.

**Acknowledgements**: We would like to thank the past and present students and postdocs at the University of Western Australia and ANSTO involved in the research work on hydrogen sensing as well as our colleagues for invaluable discussions.